\documentclass[fleqn,usenatbib]{mnras}
\newcommand{\edit}[1]{#1}
\newcommand{\edittwo}[1]{#1}

\usepackage{newtxtext,newtxmath}
\usepackage{physics}
\usepackage{booktabs}
\usepackage{xcolor}
\usepackage{multirow}
\usepackage{natbib}
\usepackage{float}

\usepackage[T1]{fontenc}

\DeclareRobustCommand{\VAN}[3]{#2}
\let\VANthebibliography\thebibliography
\def\thebibliography{\DeclareRobustCommand{\VAN}[3]{##3}\VANthebibliography}

\usepackage{graphicx}	
\usepackage{amsmath}

\usepackage{caption}
\usepackage[skip=0.5ex]{subcaption}

\title[YSO Variability]{Towards an understanding of YSO variability: A multi-wavelength analysis of bursting, dipping, and symmetrically varying light curves of disc-bearing YSOs.}

\author[Lakeland \& Naylor]{
Ben S. Lakeland\thanks{E-mail: bsl204@exeter.ac.uk} and
Tim Naylor\\
Department of Physics and Astronomy, University of Exeter, Exeter, EX4 4QL, UK\\
}

\date{Accepted XXX. Received YYY; in original form ZZZ}

\pubyear{2022}

\begin{document}
\label{firstpage}
\pagerange{\pageref{firstpage}--\pageref{lastpage}}
\maketitle

\begin{abstract}
Using simultaneous optical and infrared light curves of disc-bearing young stars in NGC 2264, we perform the first multi-wavelength structure function study of YSOs. \edit{We} find that dippers \edittwo{have larger variability amplitudes} than bursters and symmetric variables at all timescales longer than a few hours. By analysing optical-infrared colour time-series, we also find that the variability in the bursters is systematically less chromatic at all timescales than the other variability types. \edittwo{We} propose a model of YSO variability in which symmetric, bursting, and dipping behaviour is observed in systems viewed at low, intermediate, and high inclinations, respectively. We argue that the relatively short thermal timescale for the disc can explain the fact that the infrared light curves for bursters are more symmetric than their optical counterparts, as the disc reprocesses the light from all rotational phases. From this model, we find that the accretion variability onto these YSOs roughly follows a random-walk.
\end{abstract}


\section{Introduction}
The photometric variability exhibited by young stellar objects (YSOs) is well established as a key identifying feature \cite[e.g.,][]{1945ApJ...102..168J}. For YSOs with circumstellar discs, this variability can arise from processes occurring around the stellar photosphere, or from within the disc. Such varied phenomena can give rise to a variety of light curve morphologies. For example, for some disc bearing YSOs, line-of-sight extinction of the central object by the circumstellar material may manifest itself as short duration dips in the flux. As well as these so-called \lq dippers\rq, there are objects whose light curves contain several abrupt increases in flux as a result of an enhanced accretion rate (\lq bursters\rq). Sections \ref{sec:prev_class} and \ref{sec:current_model} describes these classes in more detail.\\

By taking advantage of the unparalleled precision and cadence offered by space-based observatories, recent photometric surveys have been able to examine the light curves of YSOs to an unprecedented level of detail \citep[e.g.,][]{morales-calderon, C14, Cody2018}. Until recently, however, much of the photometric analysis of YSOs has relied on period searches, e.g. via a periodogram \citep{Lomb, Scargle}. Whilst this has been fruitful, only around 30-70 per cent of monitored YSOs are observed to display periodic or quasi-periodic variations \citep[e.g.,][]{Herbst, Littlefair2010, Venuti17}. Relying too heavily on periodicity searches therefore ignores between one third and two thirds of the disc bearing YSOs. Recent studies \citep{S20, Venuti21} tackle this limitation by employing a structure function analysis of YSOs in Cep OB3b and the Lagoon Nebula, respectively. Their analyses have demonstrated that the structure function is a powerful tool to analyse the behaviour of both periodic and aperiodic time series. The difference between this approach and traditional variability analyses can be summarised by noting that the structure function identifies a timescale rather than a \edit{coherent} period.\\

In this paper, we use the \textit{CoRoT} and \textit{Spitzer} light curves provided as part of the coordinated synoptic investigation of the NGC 2264 star forming region \citep[see Section \ref{sec: data}]{C14}. We use the morphological classifications provided by \citet{C14}. In particular, we focus on objects with \textit{CoRoT} light curves which demonstrate bursts, dips, and symmetric behaviour (Section \ref{sec:prev_class}). For each light curve, we produce a normalised magnitude distribution (Section \ref{sec: normed_mag}) which shows whether an object is more likely to be found in a brighter or fainter state. Following on from this, we take advantage of the simultaneous optical and infrared light curves to produce single-wavelength (Section \ref{sec:sf}) and colour (Section \ref{sec: colour_sf}) structure functions.
It is worth noting that the normalised magnitude distributions and structure functions offer complementary analyses. \edit{As such, we provide discussion of the analyses from Sections \ref{sec: normed_mag}, \ref{sec:sf}, and \ref{sec: colour_sf}, as well as an overview of the current intepretation of YSO variability, in Section \ref{sec: discussion}}

We find results that both are consistent with the current picture of variability of disc-bearing YSOs, and motivate a model of variability with a strong dependence on viewing inclination. Finally, we conclude in Section \ref{sec: conclusions}.\\

\section{Data and Region} \label{sec: data}
 NGC 2264 is a young galactic open cluster located in the Mon OB1 association. Located at an approximate distance of $\approx 700-800$ pc \citep{Park2000}, its rich selection of pre-main-sequence (PMS) objects and relative lack of foreground extinction have made it an attractive target for variability studies \citep[e.g.,][]{Alencar2010, Rebull14}. Its estimated age of $\approx$ 3 Myr \citep{Dahm08} places NGC 2264 between the sample of $\approx$ 0.7 Myr-old PMS objects in the Lagoon Nebula \citep{Prisinzano} and the $\approx$ 6 Myr-old Cep OB3b association \citep{Bell13} previously studied with structure functions.\\

The data used in this paper are taken from the Coordinated Synoptic Investigation of NGC 2264 (CSI 2264) undertaken by \citet{C14}.
As part of that survey, the authors assign each object a unique identifier (\lq Mon-XXX\rq) which we will use in place of observatory specific identifiers.
A detailed description of data collection and reduction are provided by \citet{C14}, which we here summarise. 

\subsection{\edit{CoRoT}} \label{sec: data_corot}

The optical data used in our analysis were obtained during the fifth short (\textit{SRa05}) run of the \textbf{Co}nvection, \textbf{Ro}tation, and planetary \textbf{T}ransits \citep[\textit{CoRoT},][]{Baglin} mission \footnote{There are also some optical light curves from observation run \textit{SRa01}. We choose to omit these data since they lack the simultaneous infrared light curves from which the \textit{SRa05} data benefit.}. \textit{SRa05} ran from 2011 December 01 to 2012 January 09 and monitored 4235 stars. For this paper, we use the light curves supplied by the CSI 2264 team\footnote{https://irsa.ipac.caltech.edu/data/SPITZER/CSI2264/ \label{fnte: data}}. These differ from the light curves provided directly from \textit{CoRoT} in the following ways.
\begin{itemize}
    \item The \textsc{WHITEFLUX} photometry has been converted to \textit{R}-band magnitudes by calibrating the logarithmic mean flux to \textit{R-}band photometry in the literature  \citep{Rebull2002, Lamm, Sung}.
    \item The light curves have been corrected for known systematic effects, most notably the \lq hot pixels\rq\  which appear in the light curves as sudden discontinuities.
    \item Some light curves with significant systematics have been entirely omitted from the sample.
\end{itemize}

The exposure time for most targets was 512 seconds, with a subset of stars with evidence of eclipses or transits being observed with a 32 second cadence mode. This high cadence mode was mainly used for stars with an \textit{R} magnitude brighter than 14$^\text{th}$. For consistency, the light curves for all of the stars are binned to a 512 second cadence by \citet{C14}.

\subsection{Spitzer} \label{sec:irac_data}

NGC 2264 was also observed by the \textit{Spitzer} \citep{Werner} \textbf{I}nfra\textbf{r}ed \textbf{A}rray \textbf{C}amera \citep[IRAC][]{Fazio} simultaneously to \textit{SRa05} (2011 December 3 to 2012 January 1). Observations were taken in the $3.6\upmu$m and $4.5\upmu$m channels (IRAC1 and IRAC2, respectively).\\

Section 3.2 of \citet{C14} details the light curve production from the basic calibrated data images from the \textit{Spitzer} Science Centre. The extracted light curves are also made available by the CSI 2264 team\footnotemark[2\label{fnte: data}]. The resulting light curves have $\approx$ 12 second exposures, separated by $\approx$ 101 minutes due to the relatively small field of view of the \textit{IRAC} instrument.
\subsection{CSI 2264 sample selection}

The final sample of \citet{C14} contains 162 disc-bearing YSOs, of which 150 have simultaneous light curves from both \textit{CoRoT} and at least one IRAC band. 140 of these sources have simultaneous light curves from both infrared bands available.
Since we wish to investigate the simultaneous multi-wavelength behaviour shown by the objects in our sample, we only retain the 150 objects with at least one infrared light curve. For details on membership criteria, the reader is directed to the appendix of \citet{C14}. In addition to the light curves detailed in Sections \ref{sec: data_corot} and \ref{sec:irac_data}, we also have spectral type information available for many of the objects \citep{Walker, Makidon, Dahm+Simon}.\\

\section{Variability classification} \label{sec:prev_class}

\label{sec: ex_classes}
\edit{\citet{C14} built upon long-standing classification schemes \citep[e.g., ][]{1954TrSht..25....1P, 1994AJ....108.1906H} by classifying each light curve into one of the following variability types: aperiodic dippers, aperiodic bursters, stochastic, quasi-periodic dippers, quasi-periodic bursters, quasi-periodic symmetric, long timescale variables, eclipsing binaries, periodic variables, multi-periodic variables, or non-variable stars. Objects were initially assigned a class based on a visual inspection of the light curves, and then used to derive a statistical classification by calculating asymmetry and quasi-periodicity statistics to summarise each light curve (Section \ref{sec: qm}). These classifications were done separately for the optical and infrared data. It is worth noting here that in this paper we will not employ the infrared morphology classifications and instead divide the YSOs by their optical behaviour. Optical morphological classifications are based on the \textit{SRa05} data (see Section \ref{sec: data_corot}). This classification scheme was expanded with a study of the Upper Scorpius and $\rho$ Ophiuchus \citep{Cody2018}, as part of the second campaign of the \textit{K2} \citep{Howell} mission with the re-purposed \textit{Kepler} spacecraft \citep{Borucki}.}\\ 

This study will focus on the YSOs in NGC 2264 identified by \citet{C14} as having bursting, dipping, or symmetrically varying optical light curves. We briefly summarise these variability classes here. Fig. \ref{fig:ex_lcs} shows example light curves from each class as well as the corresponding structure function for each waveband (see Section \ref{sec:sf} for more details).\\

From the 150 objects with simultaneous \textit{CoRoT} and \textit{Spitzer} photometry (see Section \ref{sec: data}), we compile a sample of 94 YSOs with optical light curves displaying bursts, dips, or symmetric variations. The morphological classes are described below. The numbers of bursters, dippers, and symmetric variables used in this paper are 21, 35, and 38, respectively. The objects not used in this paper were identified by \citet{C14} as having either purely periodic (or multi-periodic) variations, non-variable light curves, or light curves which are unable to be classified. We note that, unlike \citet{C14}, we group the aperiodic and quasi-periodic variables together into the classes outlined in this section. Since we are interested in the simultaneous optical and infrared variability, we do not include objects for which \cite{C14} do not provide \textit{Spitzer} and \textit{CoRoT} photometry. The choice to limit our investigation to the morphology classes outlined in this section means that we retain 63 per cent of the entire sample, and 85 per cent of the objects not identified by \citet{C14} as non-variable, or unclassifiable from the optical light curves.

\begin{figure*}
    \centering
    \includegraphics[width=\textwidth]{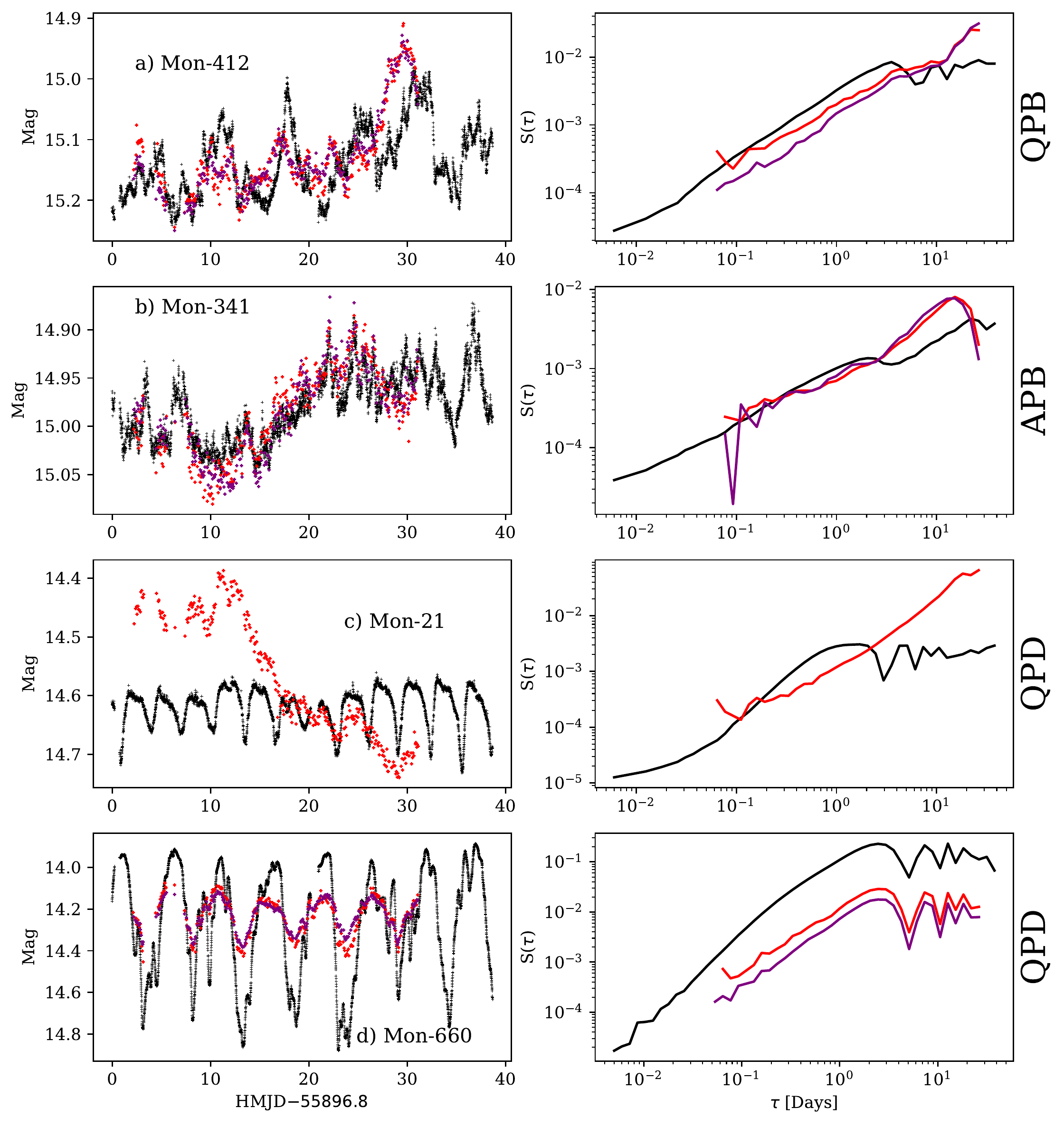}
    \caption{(Left) \textit{CoRoT} \textit{R}-band (black), IRAC [3.6] (red), and IRAC [4.5] (purple) light curves for some example YSOs. The infrared light curves have been vertically displaced so that the optical and infrared magnitudes have the same median. We select light curves from the following classes to illustrate some types of behaviour: quasi-periodic burster (panel \textit{a}), aperiodic burster (panel \textit{b}), quasi-periodic dipper (panels \textit{c} and \textit{d}), aperiodic dipper (panel \textit{e}), quasi-periodic symmetric (panels \textit{f} and \textit{g}), and stochastic (panel \textit{h}). (Right) The corresponding structure functions for each object. See Section \ref{sec:sf} for more details.}
    \label{fig:ex_lcs}
\end{figure*}
\begin{figure*}\ContinuedFloat
    \centering
    \includegraphics[width=\textwidth]{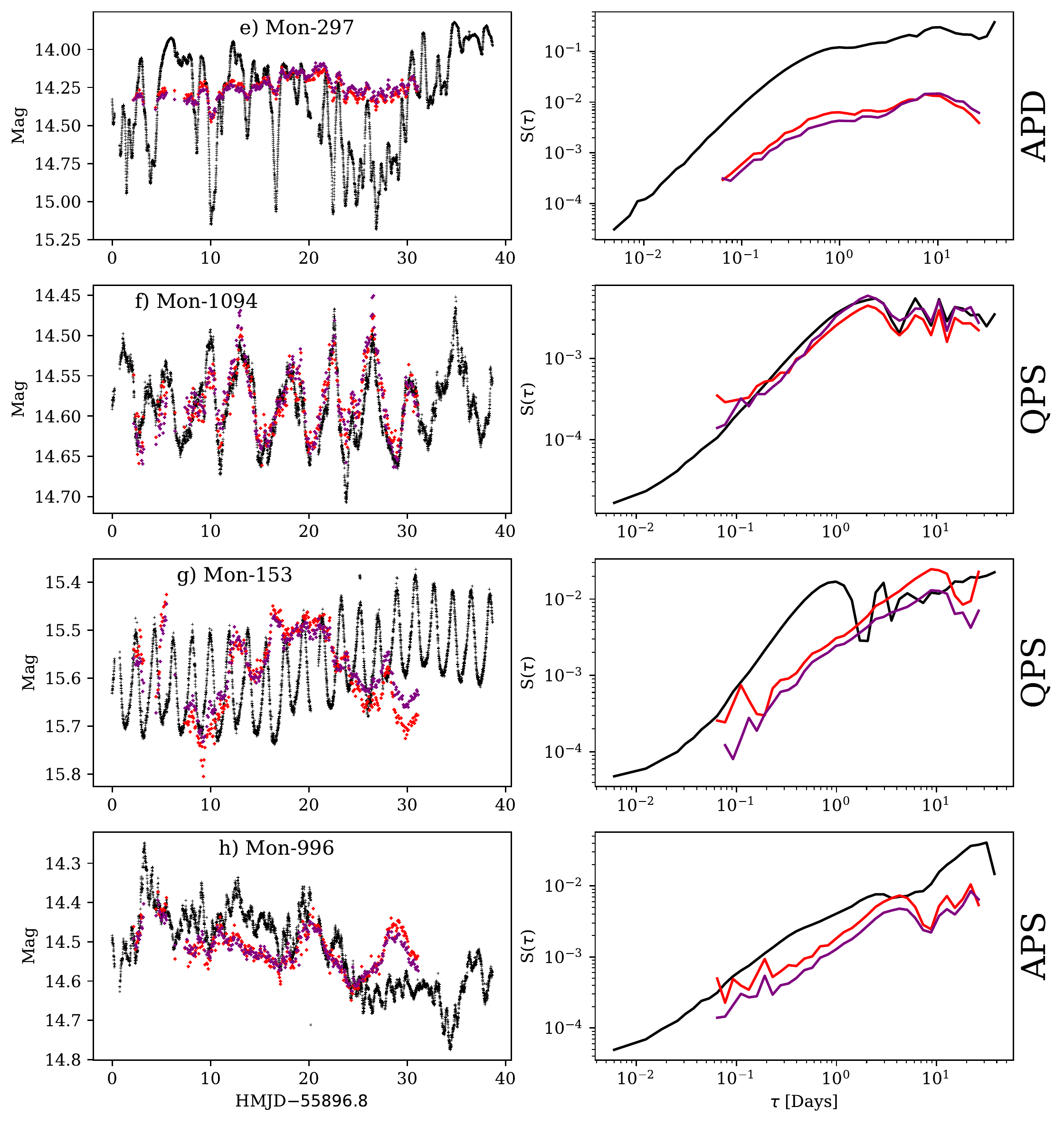}
    \caption{Continued.}
    \label{fig:ex_lcs}
\end{figure*}

\subsection{Light curve classes} \label{sec: var_classes}

\edit{In this section, we provide an observational overview of the morphological classes of light curves we study in this work. In Section \ref{sec:current_model}, we expand on this and summarise the favoured model for the physical processes behind the variability.}

\subsubsection{Bursters}

\citet{C14} define bursters as YSOs whose light curves display rapid and abrupt increases in flux. An important feature that differentiates this class from other brightening events (e.g., stellar flares, FUOr/EXOr outbursts) is that the flux returns to the quiescent value on a similar timescale to the rise (i.e., the bursts are symmetric). It is also noticeable that bursters typically display many bursts over the course of the $\sim 40$ day observations of \citet{C14}. This contrasts dramatically to the inferred recurrence timescale for large-amplitude, long-lasting bursts in Class II YSOs of 112 kyr \citep{Pena19}. We also note that the bursts observed in this sample are generally low amplitude, this is in sharp contrast to the sample of \citet{Cody17} which can show variability of up to several times the median flux. \edit{Using the Stetson index, \citet{C14} identify stars with well correlated optical and infrared light curves and find that the majority of these are identified as optical bursters or stochastic (Section \ref{sec: symmetric}) stars.}\\ 

Bursts are observed to occur aperiodically (with no well defined period), or quasi-periodically (occurring regularly, but with changes to the light curve between bursts). \citet{C14} identified 21 optically bursting light curves, and 8 objects which display bursting in their infrared light curves. Panels \textit{a} and \textit{b} in Fig. \ref{fig:ex_lcs} show examples of the burster class. The sample of bursters used in this paper consist of the 21 optically bursting YSOs identified by \citet{C14}.\\

\subsubsection{Dippers}

Dipping behaviour is a well documented phenomenon in the optical light curves of YSOs, a notable example being the prototype AA Tau \citep{Bouvier}. These light curves are characterised by significant fading events lasting up to a few days. In objects which have simultaneous optical and infrared dips, the dips are generally observed to be more extreme in the optical \citep[e.g., ][]{C14}.\\

In total, \citet{C14} found 35 optical dippers and 7 infrared dippers from the disc-bearing sample. These are a mix of short duration periodic dips \citep[panel \textit{c} in Fig. \ref{fig:ex_lcs}]{Stauffer15}, AA Tau analogues \citep[long duration periodic dips which have significant structure at their minima,][ panel \textit{d}]{McGinnis15}, or aperiodic dippers (panel \textit{e} in Fig. \ref{fig:ex_lcs}). Even though several of these objects have a detectable period, they are classified as quasi-periodic since the dip profile will change between dips. Our set of dippers consists of the objects identified as optical dippers by \citet{C14}. \\

\subsubsection{Symmetric light curves}\label{sec: symmetric}

Objects which display no clear preference for brightening or fading are labelled by \citet{C14} as \lq stochastic\rq\ for the aperiodic objects and \lq quasi-periodic symmetric\rq\ for those which display quasi-periodic variations. Those authors suggested that the aperiodic stochastic variables may exhibit complex behaviour which is a combination of bursting and dipping discussed above. The group identified from their optical light curves as quasi-periodic symmetric variables in \citet{C14} encompasses a range of behaviours. A number of the YSOs display almost sinusoidal variations, possibly attributable to features on the stellar surface rotating in and out of view. There are also light curves which display more complex behaviours which may arise from star-disc interactions. Since we are predominantly interested in the complex phenomena arising from the disc and its interaction with the central object, we exclude objects whose light curves are dominated by sinusoidal variations.\footnote{Namely, we exclude the objects with the following Mon IDs: 12, 58, 103, 153, 177, 765, and 1114} Owing to the small number of such objects, this does not appreciably change any of our results or conclusions. Panels \textit{f} and \textit{g} in Fig. \ref{fig:ex_lcs} show examples of the quasi-periodic symmetric class, with panel \textit{h} demonstrating the aperiodic stochastic class.\\

In total there are 20 aperiodic stochastic variables, and 18 quasi-periodic symmetric variables selected from the optical light curves\footnote{From the infrared light curves, there are 8 and 18 stochastic and quasi-periodic symmetric objects, respectively. As with the other variability classes, we use the optical classifications.}. In this paper, we refer to all of these objects as symmetric variables.\\

\subsection{Statistical classification} \label{sec: qm}
\citet{C14} initially classified objects by visually examining the light curves and assigning them to one of the classes mentioned above. Additionally, they calculate two statistics to quantitatively analyse the light curves. The first is a measure of flux asymmetry, defined as
\begin{equation}
    M = \frac{\left < d_{10}\right> - d_\text{med}}{\sigma},
\end{equation}
where $\left < d_{10}\right>$ is the mean of the upper and lower 5\% of the data, $d_\text{med}$ is the median of all the data, and $\sigma$ is the RMS. Since the data supplied are in magnitudes, a large positive $M$ corresponds to dipping behaviour, and a large negative value to bursting. The second statistic is a measure of quasi-periodicity, defined as 
\begin{equation}
    Q = \frac{\text{RMS}_\text{resid}^2 - \sigma^2}{\text{RMS}_\text{raw}^2 - \sigma^2},
\end{equation}
where $\text{RMS}_\text{raw}$ and $\text{RMS}_\text{resid}$ are the RMS of the raw light curve and a phase subtracted light curve, respectively, and $\sigma$ is the uncertainty. The phase subtracted light curve is calculated by folding the raw light curve on the dominant period identified by the auto-correlation function and subtracting this phase trend from the raw light curve. For purely periodic variables, the phase subtracted curve would consist only of noise, whereas the more quasi-periodic and aperiodic objects would demonstrate structure in their phase subtracted light curves.\\

The YSOs in the CSI 2264 study were placed on a $Q$-$M$ plot \citep[fig. 31,][]{C14} and used to derive classification boundaries which were then applied to the YSOs in $\rho$ Oph/Upper Sco \citep{Cody2018}.

\section{Normalised magnitude distributions} \label{sec: normed_mag}

\begin{figure*}
    \centering
    \includegraphics[width = \textwidth]{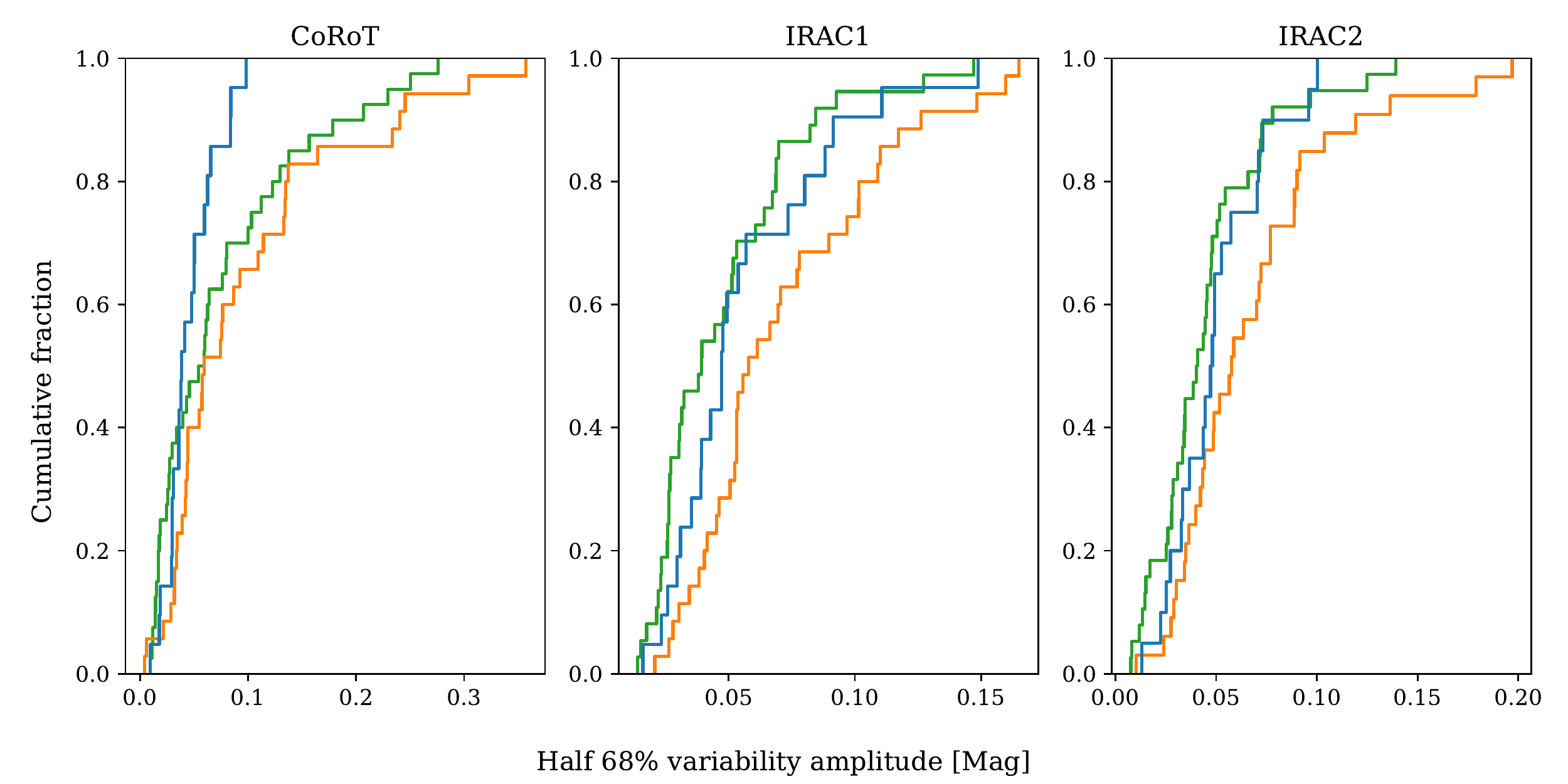}
    \caption{Cumulative distributions of half 16-84 inter-percentile range for bursters (blue), dippers (orange), and symmetric stars (green).}
    \label{fig:ah68_cumulative}
\end{figure*}

\begin{figure*}
    \centering
    \includegraphics[width=\textwidth]{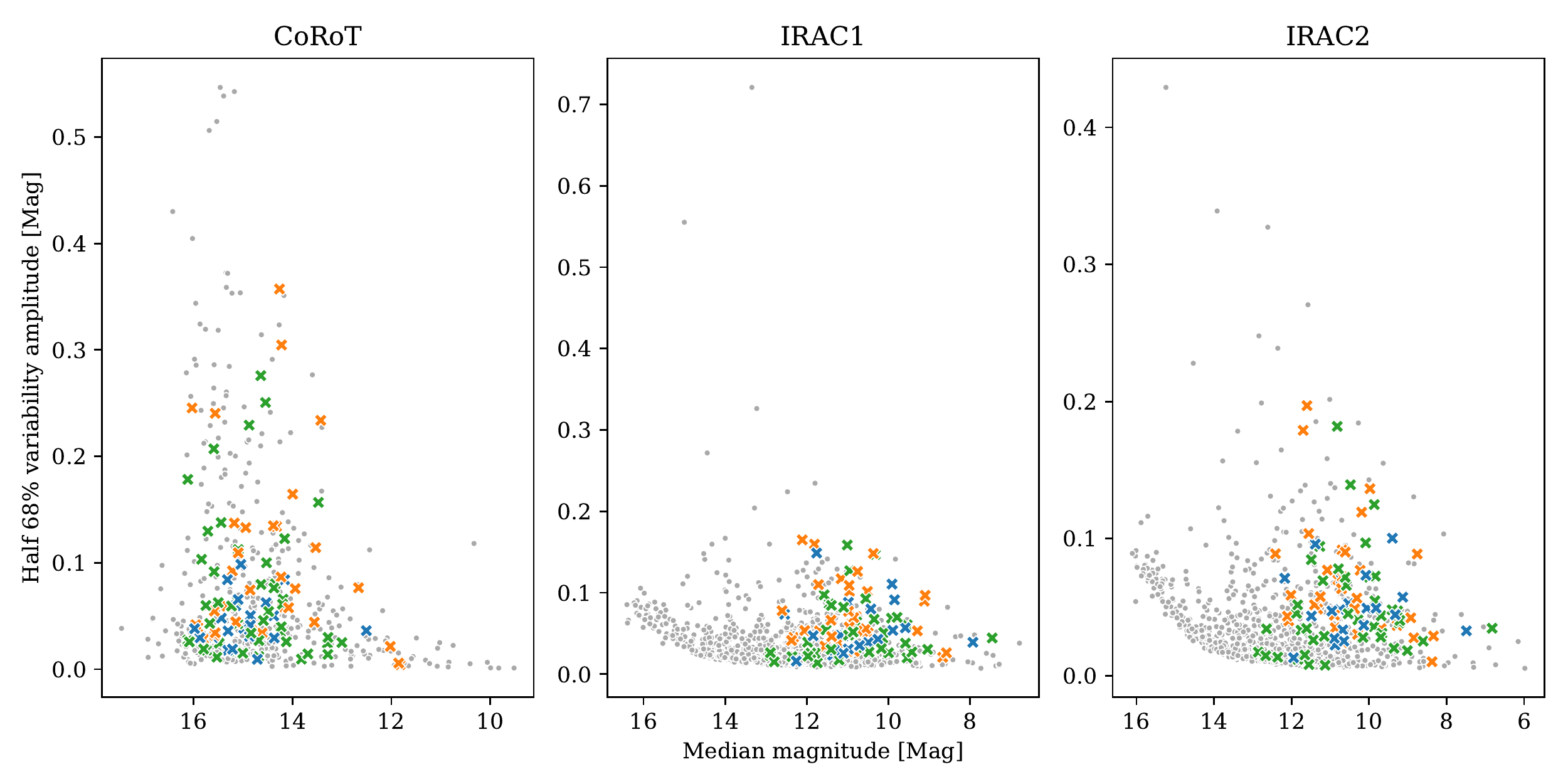}
    \caption{Median magnitude vs half 16-84 inter-percentile range for bursters (blue), dippers (orange), and symmetric stars (green). Grey points represent \edit{all objects in the greater NGC 2264 region observed in the respective waveband.}}
    \label{fig:ah68_medmag}
\end{figure*}

\begin{figure*}
    \centering
    \includegraphics[width=\textwidth]{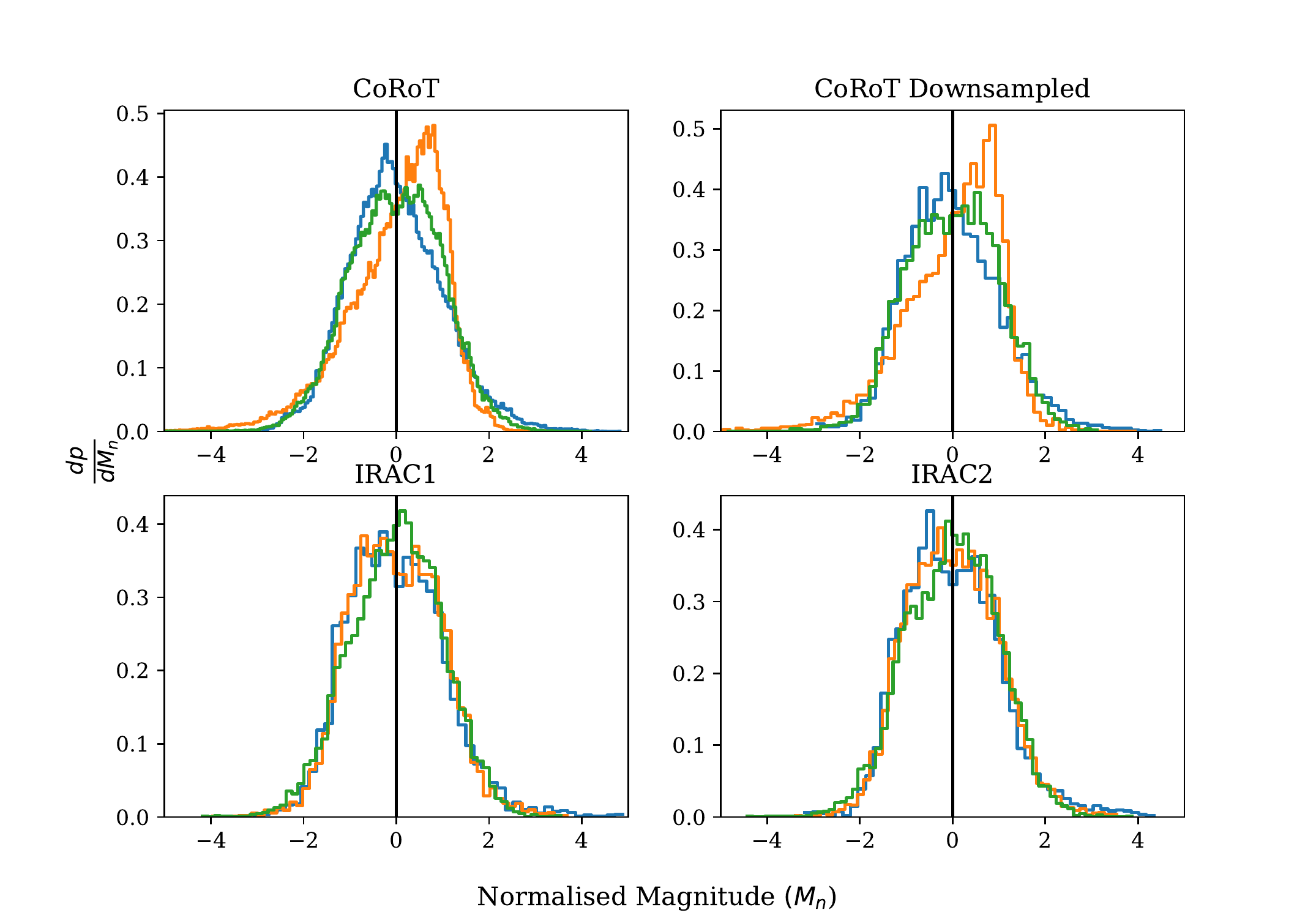}
    \caption{From top-left to bottom-right, the normalised magnitude distributions for the full \textit{R-}band light curves, \textit{R-}band light curves re-sampled to the IRAC cadence, and the [3.6] and [4.5] light curves. Bursters, dippers, and symmetric variables are shown in blue, orange, and green, respectively. A \edit{positive} value of normalised magnitude indicates a YSO being observed in a brighter state than the mean for that object. A normalised magnitude of zero is shown with a black line.}
    \label{fig:normed_mag}
\end{figure*}

As a measure of variability in the light curves, we use half the 16-84 inter-percentile range \citep[see][]{S20}. For a normally distributed signal, this is equivalent to the RMS, and is an estimate of variability amplitude which is robust against outliers and reduces the dependence on the number of observations. Fig. \ref{fig:ah68_cumulative} shows the cumulative distribution function of the variability amplitude for the bursters, dippers, and symmetric variables.
The distributions of $A_{H68}$ show that the overall variability level is larger for the dippers than for the bursters, in all wavelengths.
Fig. \ref{fig:ah68_medmag} shows that this is not simply due to the objects occupying a different range of magnitudes.
From this half 16-84 inter-percentile range, $A_{H68}$, we calculate the distributions of normalised magnitudes, defined by \citet{S20} as 
\begin{equation} \label{eq: m_n}
    M_n = \frac{\left<m\right> - m}{A_{H68}},
\end{equation}
where $m$ is an individual magnitude and $\left<m\right>$ is the mean magnitude of the light curve. This is constructed in such a way as to allow us to compare magnitude distributions between objects (see Fig. \ref{fig:normed_mag}). A positive value of normalised magnitude indicates a star being in a brighter state than its mean level and vice versa.\\

\edit{This is related to the $M$ asymmetry metric devised by \citet{C14}. We find a very strong anti-correlation between $M$ and the Fisher skew of $M_n$ for each light curve\footnote{\edit{The anti-correlation arises from the negative sign of the magnitude in Eq. \ref{eq: m_n}}}. This is unsurprising when one considers that the definition of $M$ is a modified skew calculation. We choose to investigate the full $M_n$ distribution as it allows for a more general investigation into the shape of the light curve distribution.\\}

To compare the classes identified by \citet{C14}, we combine the distributions of $M_n$ for each source, divided by optical morphology\footnote{When combining the distributions, we normalise the histogram to the total number of observations for each morphological class (i.e., despite the burster class being smaller than the other classes, and thus having fewer observations, the normalised magnitude distributions for all morphological classes have the same area).}. The top left panel of Fig. \ref{fig:normed_mag} shows that the bursters and dippers have rather different distributions of $M_n$. Although both are asymmetric (which is unsurprising when one considers that these are objects chosen for their light curve asymmetry), the bursters have a significant bright and faint tail, whereas the dippers have a very sharp drop-off for a normalised magnitude greater than about two. The symmetric group predictably has a more symmetric distribution than either the bursters or dippers, but the distribution is generally more similar to the distribution for the bursters.\\

The lower two panels of Fig. \ref{fig:normed_mag} show the same distribution for the 3.6$\upmu$m and 4.5$\upmu$m data, respectively. We group the light curves by their optical morphology, with the same colours as in the first panel. In both distributions, the optical burster Mon-804 has been removed since the infrared light curves demonstrate anomalous behaviour which skews the study of the bursters as a whole (see Appendix \ref{sec: mon-804}). We can see that both the dippers and bursters have normalised magnitude distributions comparable to the symmetric class. To \edit{verify that this is not an effect of} the different sampling \edit{of the light curves}, we resampled the \textit{R-}band light curves to the IRAC cadence. The top right panel of Fig. \ref{fig:normed_mag} shows the resulting normalised magnitude distribution. The distributions for the dippers and bursters keep their characteristic shapes after downsampling, demonstrating that the difference between the optical and infrared distributions is a \edit{genuine} effect of wavelength, not  simply a difference in sampling. \edit{For clarity, any further references in this work to the \textit{CoRoT} data are to the full, non-downsampled, data.} We will discuss these results along with those from Sections \ref{sec:sf} and \ref{sec: colour_sf} in Section \ref{sec: discussion}.

\section{Structure functions} \label{sec:sf}

Whilst the normalised magnitude distributions detailed in the previous section demonstrate the range of magnitudes YSOs may explore, it does not tell us anything about the timescale of this variability. For this we wish to analyse the correlation within the light curve. To this end, we follow the prescriptions of \citet{S20} and \citet{Venuti21} and use the structure function, since many of the objects in our sample show aperiodic variability. \edit{Other studies which have used similar analyses to investigate aperiodic behaviour of YSOs include the \lq pooled sigma\rq\ \citep{2017MNRAS.465.3889R} and $\Delta$m-$\Delta$t \citep{2015ApJ...798...89F} studies, however we aim to compare results with \citet{S20} and \citet{Venuti21}, and so opt for the structure function.} \\

 Whilst the structure function is a longstanding analysis tool in the field of extra-galactic astrophysics \citep[e.g., ][]{Simonetti, Hughes, deVries}, its use in the study of YSOs is relatively new. The first \edit{extensive} study of YSOs with structure functions was undertaken by \citet{S20} who used $i$-band photometry of over 800 YSOs in the Cep OB3b association at timescales ranging from one minute to ten years. The choice of the structure function as an analysis tool was motivated in part by their unevenly sampled observations, and the need to be able to analyse aperiodic signals. Amongst their results, they found that the $i$-band variability at all timescales is systematically lower for later YSO evolutionary class. In addition, the authors found that the observed structure functions for Class IIs could be explained if all Class IIs exhibit a roughly power-law variability spectrum driven by accretion rate changes with some Class IIs exhibiting an additional periodic component arising from the rotation of the star.\\
 
 Later, \citet{Venuti21} used structure functions to analyse YSOs in the Lagoon nebula. Unlike \citet{S20}, those authors benefit from evenly sampled data from the ninth campaign of the K2 \citep{Howell} mission. Those authors divide the sample into the variability types introduced in \citet[see Section \ref{sec: var_classes}]{C14}. \edit{We will compare our findings with those of \citet{Venuti21} in Section \ref{sec: comp_venuti}}\\
 
 The CSI 2264 data set is unique when compared to these previous studies, in having simultaneous time series in optical and infrared wavelengths. This allows us not only to examine the optical or infrared variability spectrum, but also to examine the colour variability (see Section \ref{sec: colour_sf}). Additionally, like the data set of \citet{Venuti21}, we also have information about the spectral types of the objects, with the sample covering the G to M star range \citep{Walker, Makidon, Dahm+Simon}.
 
 \subsection{Definition} \label{sec:sf_def}
 
For a light curve with magnitudes $m(t)$ observed at times $t$, we define the structure function as 
\begin{equation}
    S(\tau) = \left <\left[ m(t) - m(t+\tau) \right]^2\right > 
\end{equation}
where the average is taken over all pairs of observations separated by a timescale $\tau$. In practice, this was calculated for logarithmically spaced bins in $\tau$.
\begin{equation}
    \label{eq: sf}
    S(\tau_1, \tau_2) = \frac{1}{p(\tau_1, \tau_2)} \sum \left( m(t_1) - m(t_2)\right)^2
\end{equation}
 where the sum is taken over the $p(\tau_1, \tau_2)$ pairs of observations that satisfy
\begin{equation}
    \tau_1 < t_1 - t_2 \leq \tau_2.
\end{equation}

\edit{Throughout the entirety of this paper, we require each structure function bin to have $p(\tau_1, \tau_2) \geq 10$ to mitigate the effect of small number statistics.}\\

Note that we follow the example of \citet{Hawkins} and \citet{deVries} by calculating the structure function in magnitudes, rather than flux \citep[e.g., ][]{Simonetti, Hughes, S20}. We do this because we believe that the ratio of fluxes is a better metric of variation than the difference. That is to say that we believe a 90 per cent rise in flux is a less significant change than a 90 per cent drop in flux since the latter is a much larger fractional change. Calculating the structure function in flux would treat these two changes equally. Of course, this may be more naturally achieved by calculating the structure function in $\log\left(\text{Flux}\right)$ or $\ln\left(\text{Flux}\right)$, but we opt to use magnitudes for historical reasons. For small fractional variations in flux, Equation \ref{eq: sf} is consistent with the definition of \citet{S20}, up to a constant factor of order unity.\\

The square root of the structure function at a timescale $\tau$, $\sqrt{S(\tau)}$ is the RMS of all observations separated in time by $\tau$.
A small value of the structure function indicates that observations separated by the corresponding timescales exhibit little variation; if we know the magnitude at a time $t$, it is easier to predict the magnitude at a time $t+\tau$. Conversely a large value of the structure function implies there is little correlation between points separated by the corresponding timescale.

 A typical schematic of a structure function is shown in \edit{fig. 7 of \citet{S20}}. At short timescales, the structure function is dominated by the measurement uncertainties and point-to-point scatter of the light curve. There is then a region of increasing variability with timescale as we begin to probe the timescales for relevant physics. This tends to follow a roughly power law relation. Finally, when we are probing timescales longer than the longest timescale of the variability, the structure function will no longer be increasing, reaching a plateau. This usually corresponds to a \lq knee\rq\ in the structure function at the characteristic timescale (see Section \ref{sec: knees}).\\
 
A general feature of structure functions is  that (for most types of variability) $S(\tau_1) \gtrsim S(\tau_2)$ provided $\tau_1 > \tau_2$. A notable exception to this rule is for a purely periodic signal since observations separated by multiples of the period would have zero variability. For example, the structure function corresponding to a signal given by $f(t) = \sin\left(\omega t\right)$ is $S(\tau) = 1 - \cos\left(\omega\tau\right)$ \citep{Findeisen}. This gives rise to a characteristic shape in log-log space (see curve 1 in Fig. \ref{fig:sf_example}). Fig. 9 in \citet{S20} demonstrates the effect of finite sampling on the shape of the structure function. Table 6 in \citet{S20} compares the power law index for some well characterised noise profiles and we expand on this discussion in Appendix \ref{sec: power_idx}.

\subsection{Comparison to PeakFind}
\citet{C14} employ a conceptually similar analysis to calculate a variability timescale for aperiodic light curves. Their approach is, for a given variability amplitude $A$, to select maxima and minima in the light curve that vary by at least $A$. The corresponding timescale for this amplitude is the median time between the consecutive minima and maxima. This \lq PeakFind\rq\ algorithm can then be used to produce an amplitude-timescale (or timescale-amplitude) plot, shown in fig. 32 in \citet{C14}. This amplitude-timescale plot is then used to derive a timescale (corresponding to the 70\% of maximum amplitude).\\

\begin{figure}
    \centering
    \includegraphics[width=\linewidth]{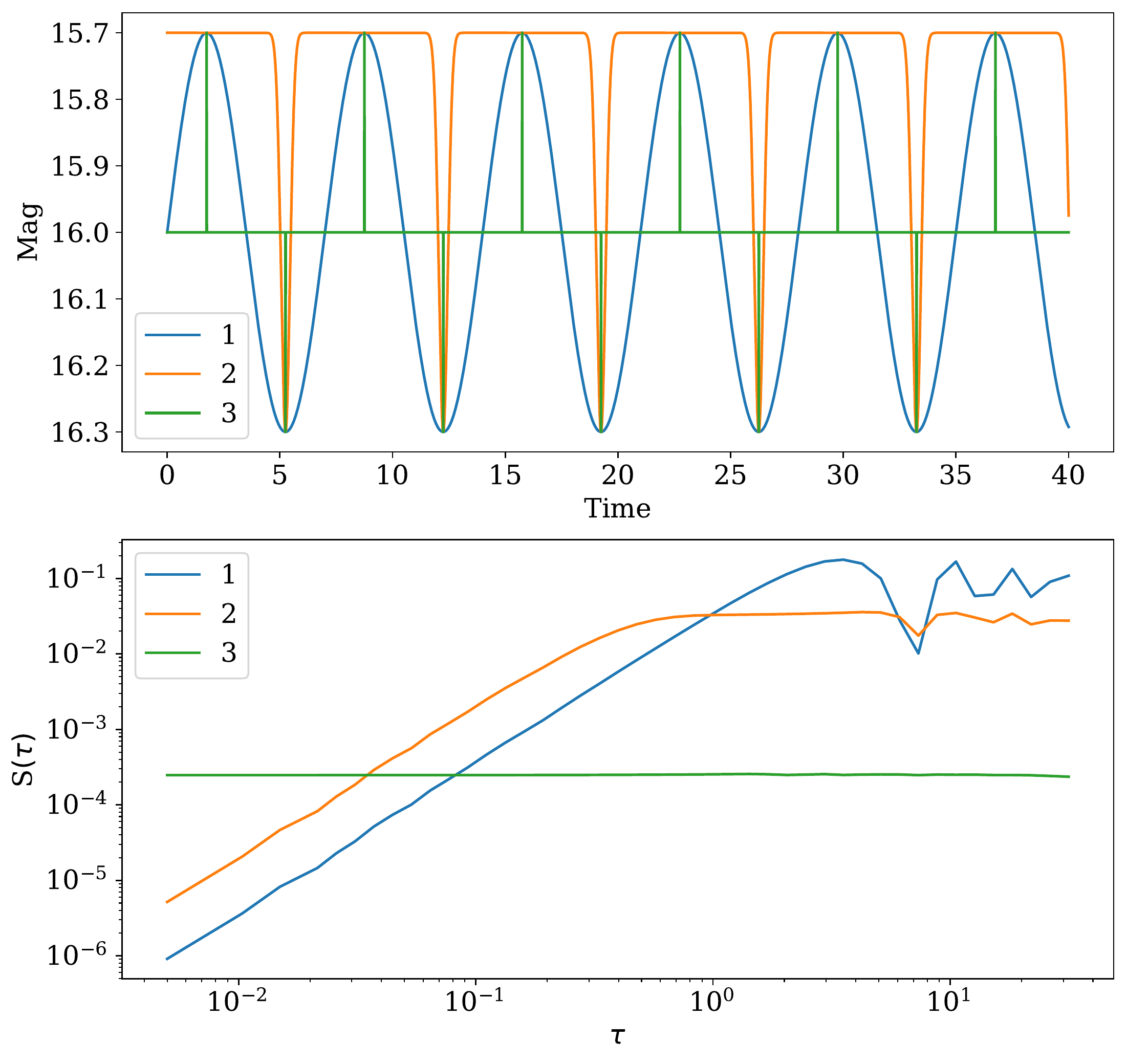}
    \caption{(Top) Three artificial light curves consisting of (1) a sinusoidal signal, (2) Gaussian dips from a constant equilibrium, and (3) alternating $\delta$-like functions. All three of these functions are constructed to have coincident minima and maxima and thus appear identical to PeakFind. (Bottom) The corresponding structure functions for the light curves in subplot a). Despite giving the same PeakFind results, the structure function makes it possible to distinguish the light curves.}
    \label{fig:sf_example}
\end{figure}

Since PeakFind only considers the relative locations of minima and maxima, any signals which have coinciding extrema will appear identical to PeakFind, regardless of the behaviour of the light curve between the extrema. This is demonstrated in Fig. \ref{fig:sf_example}. The first panel shows three basic \lq light curves\rq. We have included a sinusoid (curve 1), periodic Gaussian dips from a constant continuum (curve 2), and alternating positive and negative $\delta$-like functions\footnote{Strictly this is a top hat function with a single observations in the brighter or fainter state. } (curve 3). These are constructed to \edit{be dramatically different signals whilst having} identical minima and maxima. This ensures that PeakFind treats them identically despite their different shapes. Panel 2 demonstrates that the structure function is able to distinguish between these cases. \edit{Notably, we highlight that the \lq knee\rq\ timescales identified in curves 1 and 2 correspond to the \lq width\rq\ of the sinudoidal variation and Gaussian dips, respectively. By contrast, both structure functions show a dip at the true period of $\approx 7$ days. For a more thorough description of the relative benefits of PeakFind analysis, the reader is directed to \citet{2015ApJ...798...89F}}.\\

\subsection{Results} \label{sec: sf_results}
\begin{figure*}
    \centering
    \includegraphics[width = \textwidth]{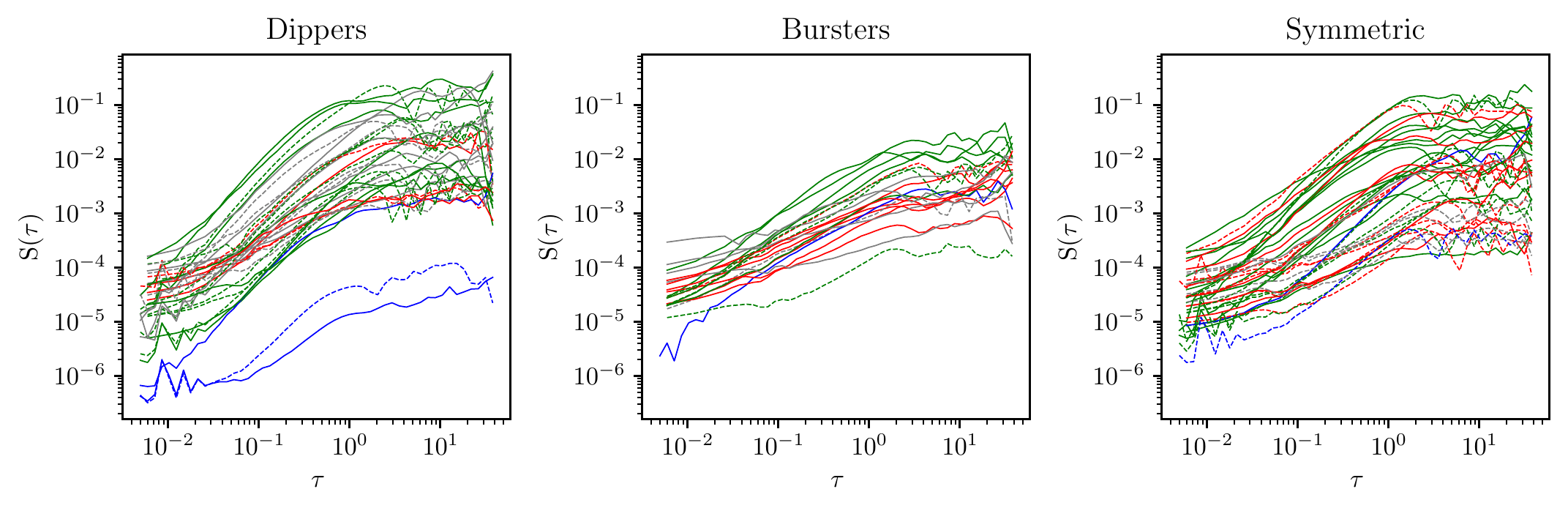}
    \caption{Structure functions of the \textit{CoRoT} light curves for our sample. G, K, and M stars are indicated by blue, green, and red lines, respectively. Grey lines indicate no spectral type information was available. Solid lines correspond to aperiodic variables, dashed lines to quasi-periodic variables. }
    \label{fig:sf_corot}
\end{figure*}

\begin{figure*}
    \centering
    \includegraphics[width = \textwidth]{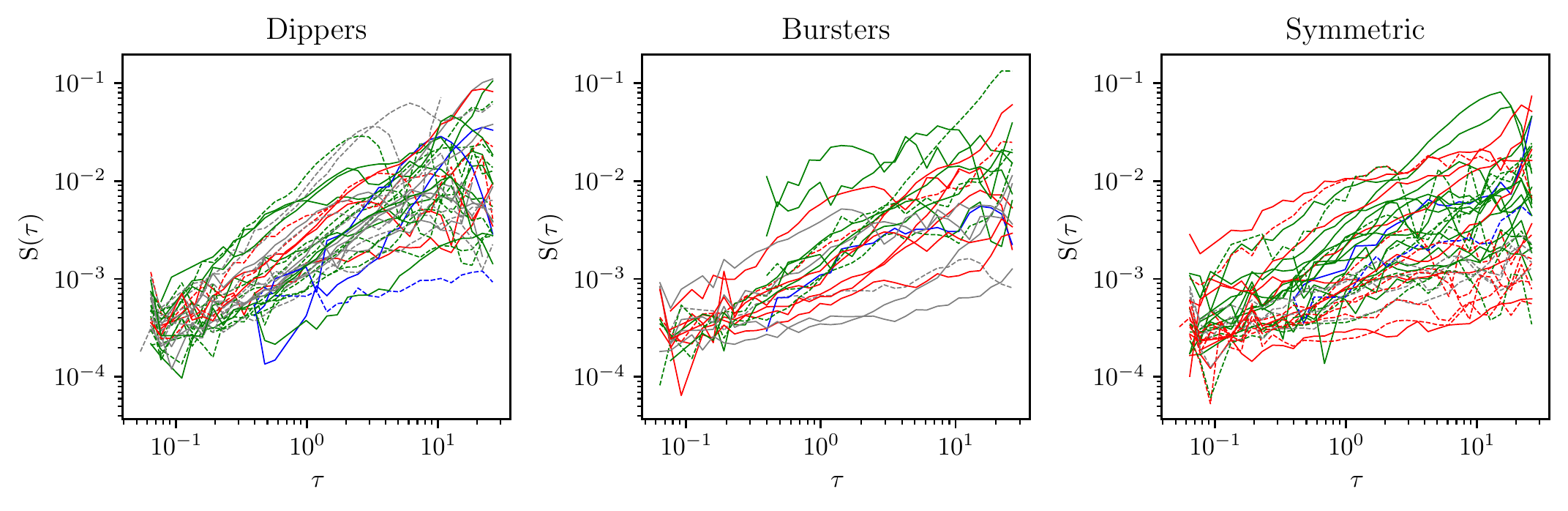}
    \caption{Structure functions of the 3.6 $\upmu$m \textit{Spitzer} light curves for our sample. G, K, and M stars are indicated by blue, green, and red lines, respectively. Grey lines indicate no spectral type information was available. Solid lines correspond to aperiodic variables, dashed lines to quasi-periodic variables. }
    \label{fig:sf_irac1}
\end{figure*}

\begin{figure*}
    \centering
    \includegraphics[width = \textwidth]{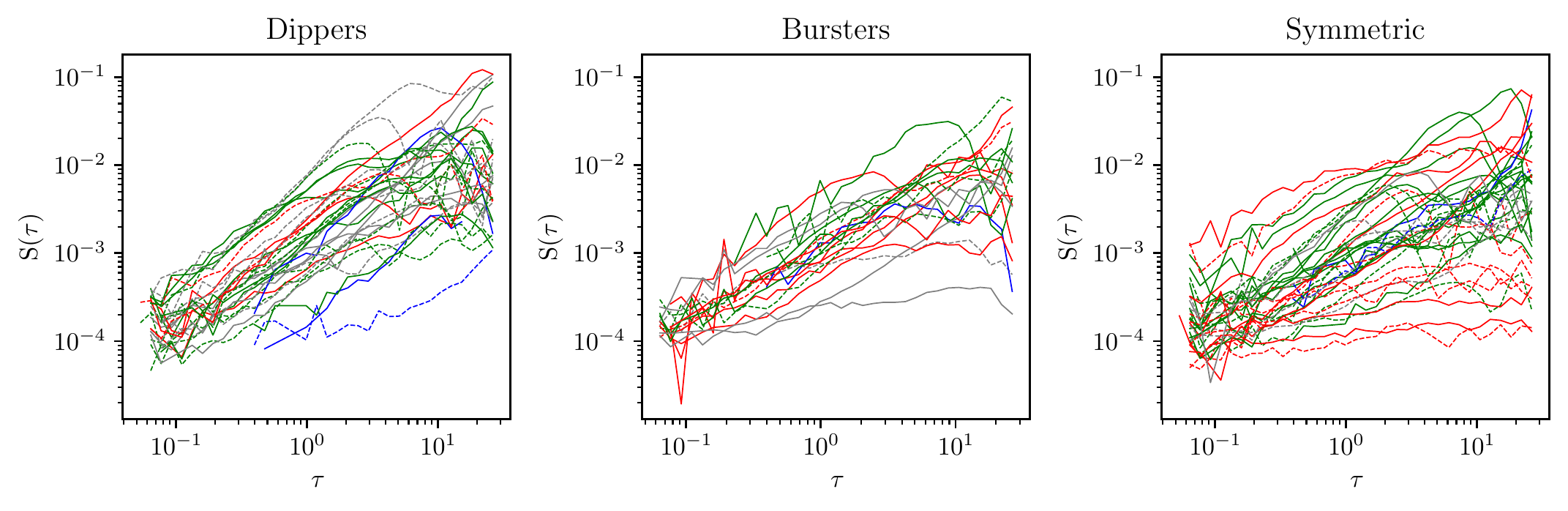}
    \caption{Structure functions of the 4.5 $\upmu$m \textit{Spitzer} light curves for our sample. G, K, and M stars are indicated by blue, green, and red lines, respectively. Grey lines indicate no spectral type information was available. Solid lines correspond to aperiodic variables, dashed lines to quasi-periodic variables. }
    \label{fig:sf_irac2}
\end{figure*}

We calculated structure functions from the \textit{CoRoT} and \textit{Spitzer} light curves for each object in our sample. Fig. \ref{fig:sf_corot} shows the structure functions calculated from the \textit{CoRoT} light curves and Figs. \ref{fig:sf_irac1} and \ref{fig:sf_irac2} show the same for the IRAC 3.6 $\upmu$m and 4.5 $\upmu$m data, respectively. As in Section \ref{sec: normed_mag}, we divide both the optical and infrared structure functions by the optical morphology. In each plot, the structure functions are colour coded by spectral type, and shown with solid or dashed lines to indicate whether the object varies aperiodically or quasi-periodically, respectively.\\

In Fig. \ref{fig:sf_corot} the dipping objects clearly exhibit stronger variability than the bursters, especially at longer timescales, \edit{in agreement with Fig. \ref{fig:ah68_cumulative}}. \edit{This indicates that the dips are generally larger amplitude than the bursts}.
\edit{The systematic difference in variability amplitudes we demonstrate here is not an artefact of the selection performed by \citet{C14} as that selection is made only on the light curve asymmetry and is therefore independent of the overall variability level. We therefore believe this to reflect a genuine difference between the variability classes.}\\

\subsubsection{Knees in the structure functions} \label{sec: knees}

We have identified structure functions with \edittwo{clear} knees in their structure functions \edittwo{(i.e., structure functions which do not show increasing variability at the longest timescales accessible from these data)}. Table \ref{tab: knee_counts_ovar} presents the results from this. It it clear that the fraction of structure functions with knees compared to those without varies between the variability types, as well as observational waveband. The majority of IRAC structure functions demonstrate variability that is still growing at longer timescales. It is worth noting that the \textit{CoRoT} light curves have a longer baseline and higher cadence, so we are naturally sensitive to a larger range of timescales in the \textit{CoRoT} structure function. Even considering this, it appears that the infrared variability generally has a longer maximum timescale. Also of note is that some stars whose structure functions appear to still be increasing at the longest timescales also show the oscillating behaviour attributed to periodic signals and there are some structure functions which demonstrate a second rise after the plateau. This corresponds to longer term, approximately linear, trends (e.g., Mon-1187) or longer period variations (e.g., Mon-525) being imposed on the continuum flux. This demonstrates that the structure function is sensitive to multiple variability modes.\\ 

\begin{table}
\centering
  \begin{tabular}{cc|c|cc|cc|cc|cc|cc|cc|}
    \hline
    & \multicolumn{3}{c|}{Single-waveband} & \multicolumn{2}{c|}{Multi-waveband}
    \\
    & \textit{CoRoT} & IRAC1 & IRAC2 & \textit{R} - [3.6] & 
    \textit{R} - [4.5]\\
    &  (\%) & (\%)& (\%) &(\%)& (\%)\\
    \hline
    Burster &  62 & 13 & 31 &  20 & 25 \\
    
    Dipper &  74 & 26 & 26 & 33 & 37  \\
    
    Symmetric &  84 & 26 & 24 & 30 & 43\\
    \\
    Aperiodic &  63 & 19 & 19 & 22 & 21 \\
    Quasi-periodic &  92 & 28 & 38 & 73 & 45 \\
    \hline
  \end{tabular}
  \caption{\edit{The percentage of stars which demonstrate a knee in their structure function. Objects are divided by \citet{C14} $M$ (i.e., into bursters, dippers, and symmetric variables) and $Q$ (i.e., into aperiodic and quasi-periodic variables) statistics. Note that in both cases, the division is done on the optical light curves. Columns indicate the waveband or colour (see Section \ref{sec: colour_sf}) the structure functions are calculated from. Objects with fewer than 150 observations in a particular light curve are omitted from the relevant count.}}
  \label{tab: knee_counts_ovar}
\end{table}

\begin{figure*}
    \centering
    \includegraphics[width=\linewidth]{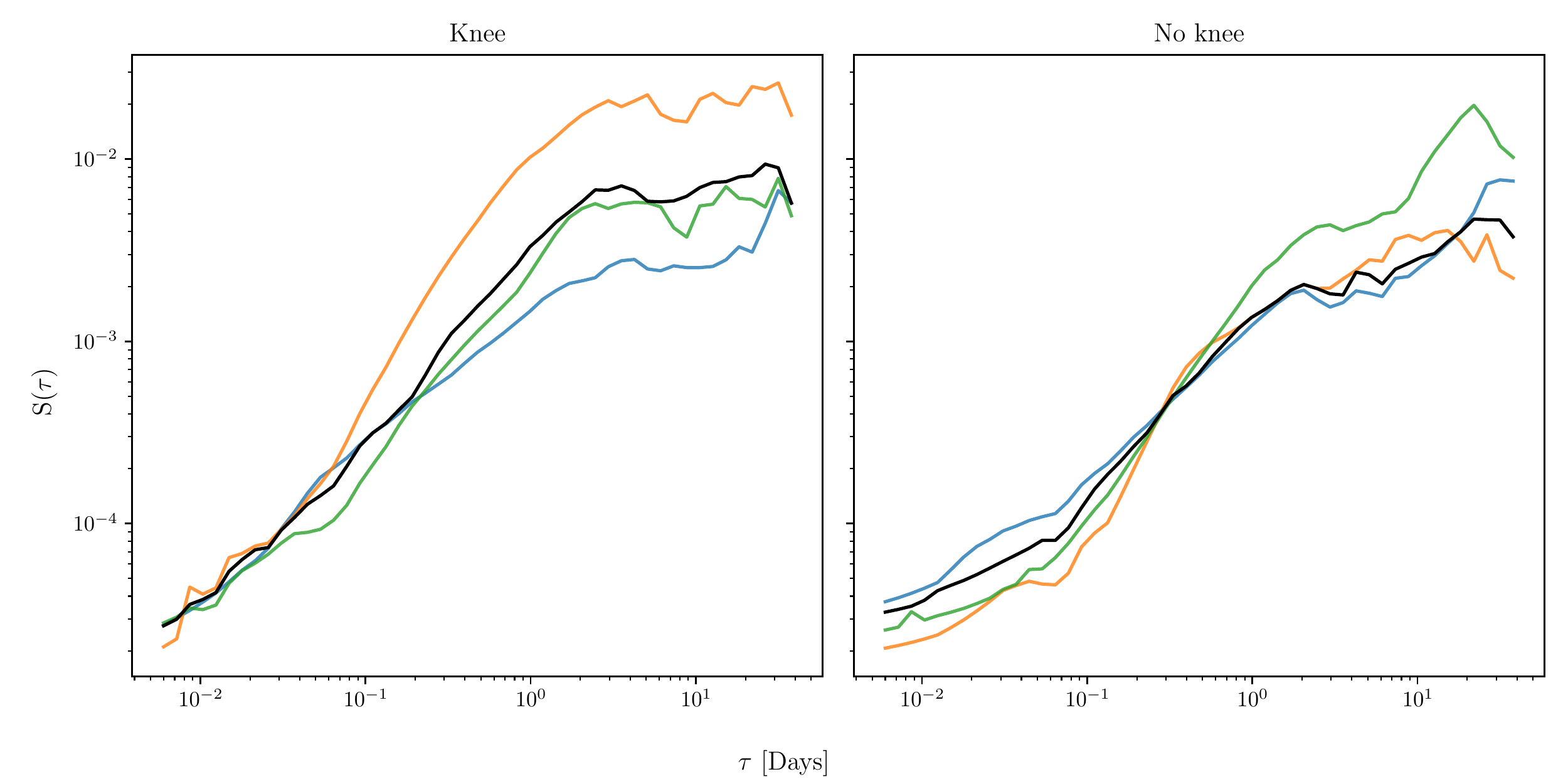}
    \caption{Median \textit{CoRoT} structure functions for bursters (blue), dippers (orange), or symmetric variables (green) divided by whether their structure functions contain a knee (left) or not (right). The black lines are the median structure functions for the entire sample. \edittwo{A number of the structure functions in the rightmost panel show knees. This is because we calculate the median value of $S(\tau)$ at each timescale, which may give rise to a knee in the median structure function despite each individual structure function not having a knee.}}
    \label{fig:med_sf_knee}
\end{figure*}

Following \citet{S20}, we divide the light curves by whether they contain a knee. Those authors find that the structure functions with knees display more variability than the structure functions without knees, despite the selection being independent of overall variability level. We do not find this to be the case for all morphology classes. Specifically, we only find a significant difference for the dippers. This can be seen clearly in Fig. \ref{fig:med_sf_knee} which shows that the median structure functions with knees and without knees are very similar for the bursters and symmetric variables\edit{, especially at longer timescales}. The black lines group the structure functions as in \citet{S20} (i.e., by presence or absence of a knee in the structure function, with no knowledge of light curve morphology). In doing so, we recover a similar result to that obtained by \citet{S20}. It is possible therefore that \citet{S20} see a difference between the amplitude of the structure functions with and without knees because their sample is dominated by dippers. 
This dominance of dippers was also suggested by \citet{S20} to explain the shape of the Class II normalised magnitude distribution. \edit{We also considered the possibility that the difference between objects with and without detectable knees were due to the lower-amplitude variables being drowned out by noise, and thus not having sufficient signal-to-noise to probe astrophysical effects. This posed two difficulties however; firstly, it is not clear why such an amplitude dependence should affect the dipper population more than the bursters and symmetric variables. Secondly, if it were intrinsically harder to identify physical timescales for the lower-amplitude variables, we might expect that the aperiodic sample of \citet{C14} to have lower overall variability than the quasi-periodic sample. Comparing the $A_{H68}$ distributions of these two samples, we find no such difference.} \\

\begin{figure}
    \centering
    \includegraphics[width=\linewidth]{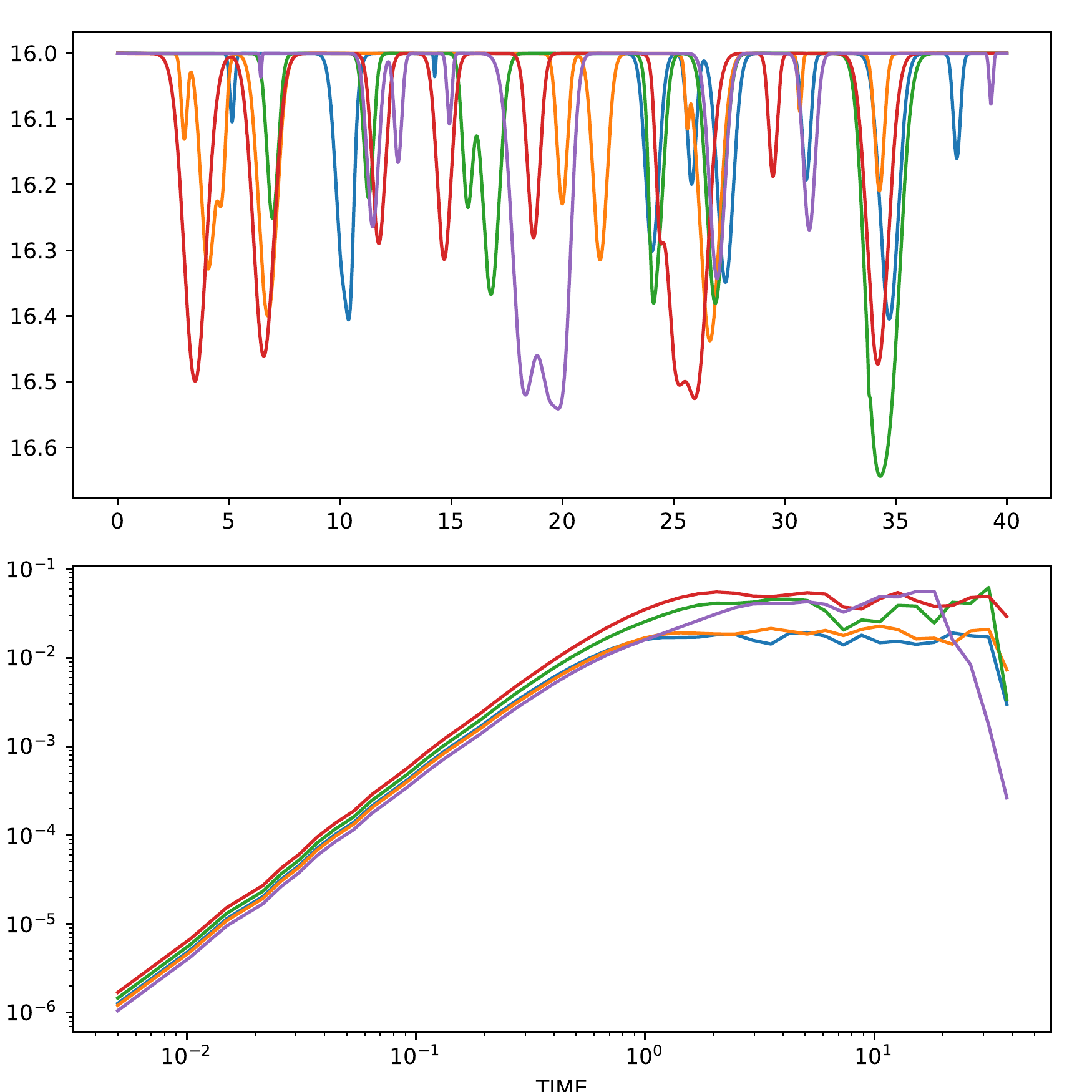}
    \caption{(Top) Five artificial light curves formed by imposing ten Gaussian dips onto a constant continuum. The location and depth of the dips are chosen at random. (Bottom) The corresponding structure functions, all demonstrating knees despite the aperiodic nature of the light curves }
    \label{fig:random_dips}
\end{figure}

It is also interesting that of the 71 YSOs whose optical structure functions have visible knees, 34 are identified as aperiodic by \citet{C14}. Similarly, a large fraction of the infrared structure functions with knees are identified as aperiodic in \citet{C14} (44 per cent for IRAC1 and 45 per cent for IRAC2).
This \edit{illustrates} that the the structure function is sensitive to the variability timescale, rather than simply identifying periods. To highlight the ability of the structure function to identify timescales in aperiodic objects, we generate five artificial light curves with Gaussian dips (Fig. \ref{fig:random_dips}). The locations and widths of the Gaussian dips are chosen at random\footnote{The depths of the Gaussian dips are fixed to be proportional to the widths.}, thus ensuring aperiodicity. Despite this, each corresponding structure function clearly has a knee. It is worth highlighting that the location of the knees are not the same in each structure function, despite the average time between dips being the same. This is to due to the light curves having dips of different durations, and thus different timescales. The difference between a structure function timescale and a period can also be seen in Fig. \ref{fig:sf_example}, where the structure functions have knees at different timescales despite being calculated from synthetic light curves that were constructed to have identical periods. Hence we \edit{highlight} that the structure function can identify timescales in signals, even when a period search may fail, but that the timescale identified is not necessarily a period.\\

\subsubsection{Timescales} \label{sec: sf_timescale}

\begin{figure}
    \centering
    \includegraphics[width=\linewidth]{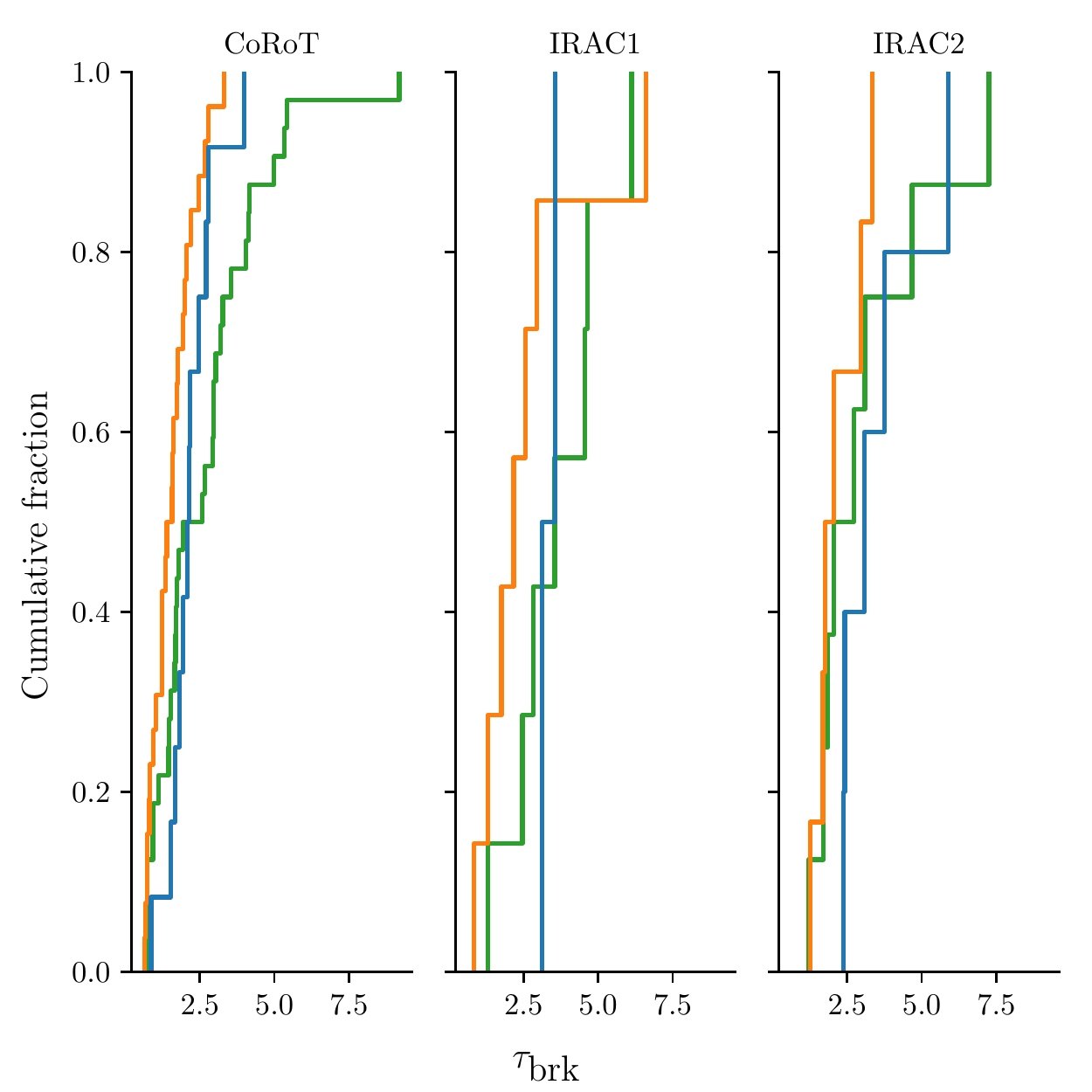}
    \caption{Distribution of timescales derived from location of knees in the structure functions. The bursters, dippers, and symmetric stars are represented by the blue, orange, and green distributions, respectively.}
    \label{fig:tbrk}
\end{figure}

\begin{figure}
    \centering
    \includegraphics[width=\linewidth]{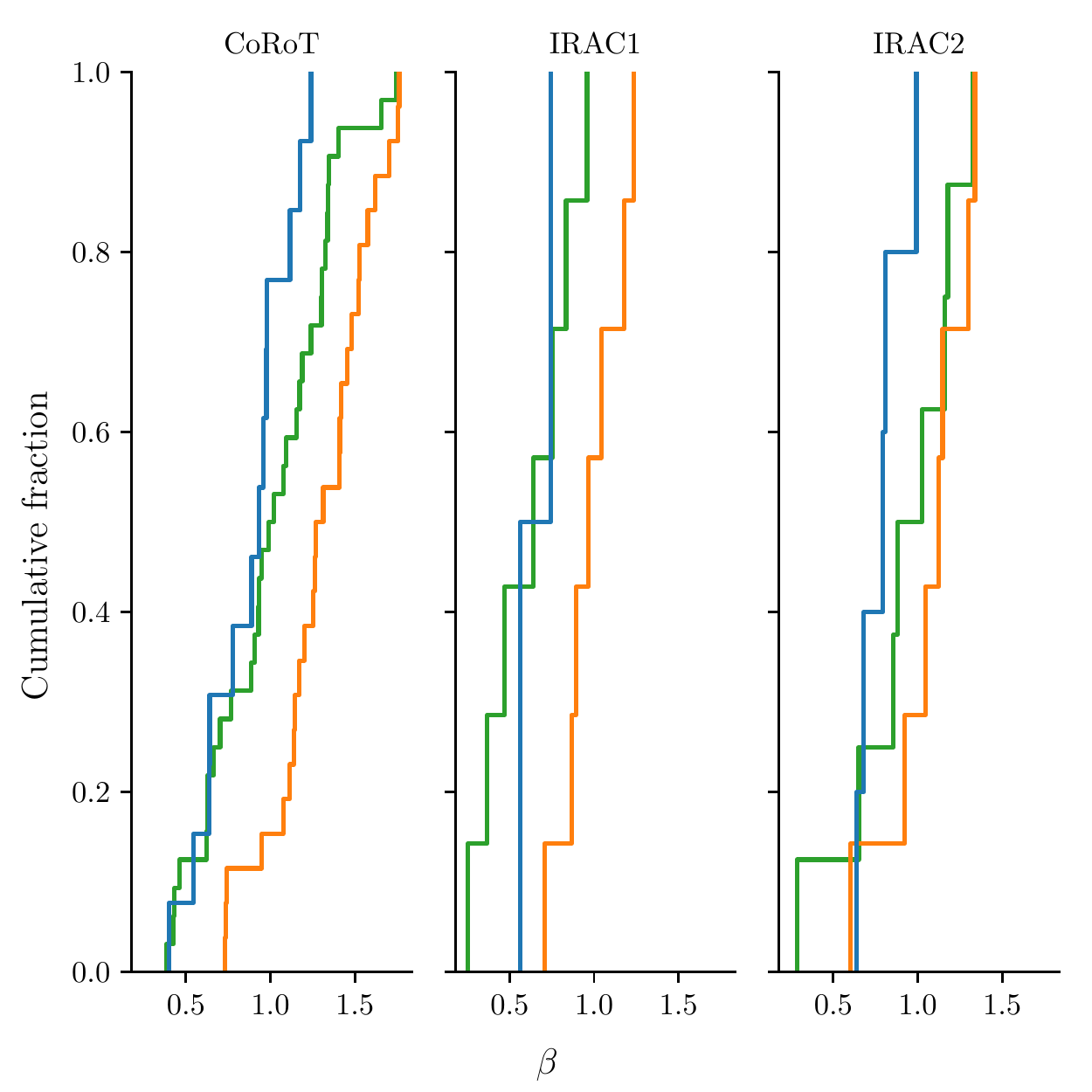}
    \caption{Distribution of structure function power law indices. The bursters, dippers, and symmetric stars are represented by the blue, orange, and green distributions, respectively.}
    \label{fig:beta}
\end{figure}

Following \citet{S20} and \citet{Venuti21}, we extract the longest timescale of variability by fitting the following power law to all structure functions with knees, 
\begin{equation}
    S(\tau) = \begin{cases}
    A \tau ^\beta & \tau \leq \tau_\text{brk}\\
    S_\text{brk} & \tau > \tau_\text{brk}
    \end{cases}
    \label{eq:fit_eq}
\end{equation}
where $A = {S_\text{brk}}/{\tau_\text{brk}^\beta}$ is defined to ensure continuity. Optimal values of $\tau_\text{brk}$, $S_\text{brk}$, and $\beta$ are found by a least squares fitting to the logarithm of Equation \ref{eq:fit_eq}. 
For YSOs without knees, we simply fit a power law relation to the structure function (i.e., the  $\tau_\text{brk} \to \infty, S_\text{brk} \to \infty$ limit of Equation \ref{eq:fit_eq}). Figs. \ref{fig:tbrk} and \ref{fig:beta} show the distributions of $\tau_\text{brk}$ and $\beta$ derived from the \textit{CoRoT} and \textit{Spitzer} structure functions, respectively. \edit{Table \ref{tab: beta_table} shows the median and RMS values of $\beta$ for the \textit{CoRoT} and \textit{Spitzer} wavelength structure functions.}  Due to the limited number of observations for some of the IRAC light curves, we only extract the above parameters from IRAC structure functions corresponding to light curves with at least 150 observations.
This leaves 69 objects for which we fit Equation \ref{eq:fit_eq} to the $3.6\upmu$m structure functions, and 76 objects for which we do the same for the $4.5\upmu$m structure functions. 
In all bands, the median value of $\tau_\text{brk}$ for the dippers is less than that for the bursters and symmetric variables, which are similar to each other. We find a median optical timescale of 2.1, 1.5, and 2.3 days for the bursters, dippers, and symmetric variables, respectively.\\ 
\begin{table}
    \centering
    \begin{tabular}{c|c|c|c}
         $\beta$ &  \textit{CoRoT} & {IRAC1} & IRAC2\\
         \hline
        Burster & 0.78 $\pm$ 0.26 & 0.65 $\pm$ 0.32 & 0.80 $\pm$ 0.27\\
        Dipper & 1.20 $\pm$ 0.36 & 0.85 $\pm$ 0.28 & 0.99 $\pm$ 0.27\\
        Symmetric & 0.94 $\pm$ 0.36 & 0.47 $\pm$ 0.28 & 0.64 $\pm$ 0.32\\
        \hline
    \end{tabular}
    \caption{Median structure function power law indices. Uncertainties given correspond to RMS values.}
    \label{tab: beta_table}
\end{table}

\begin{figure}
    \centering
    \includegraphics[width=\linewidth]{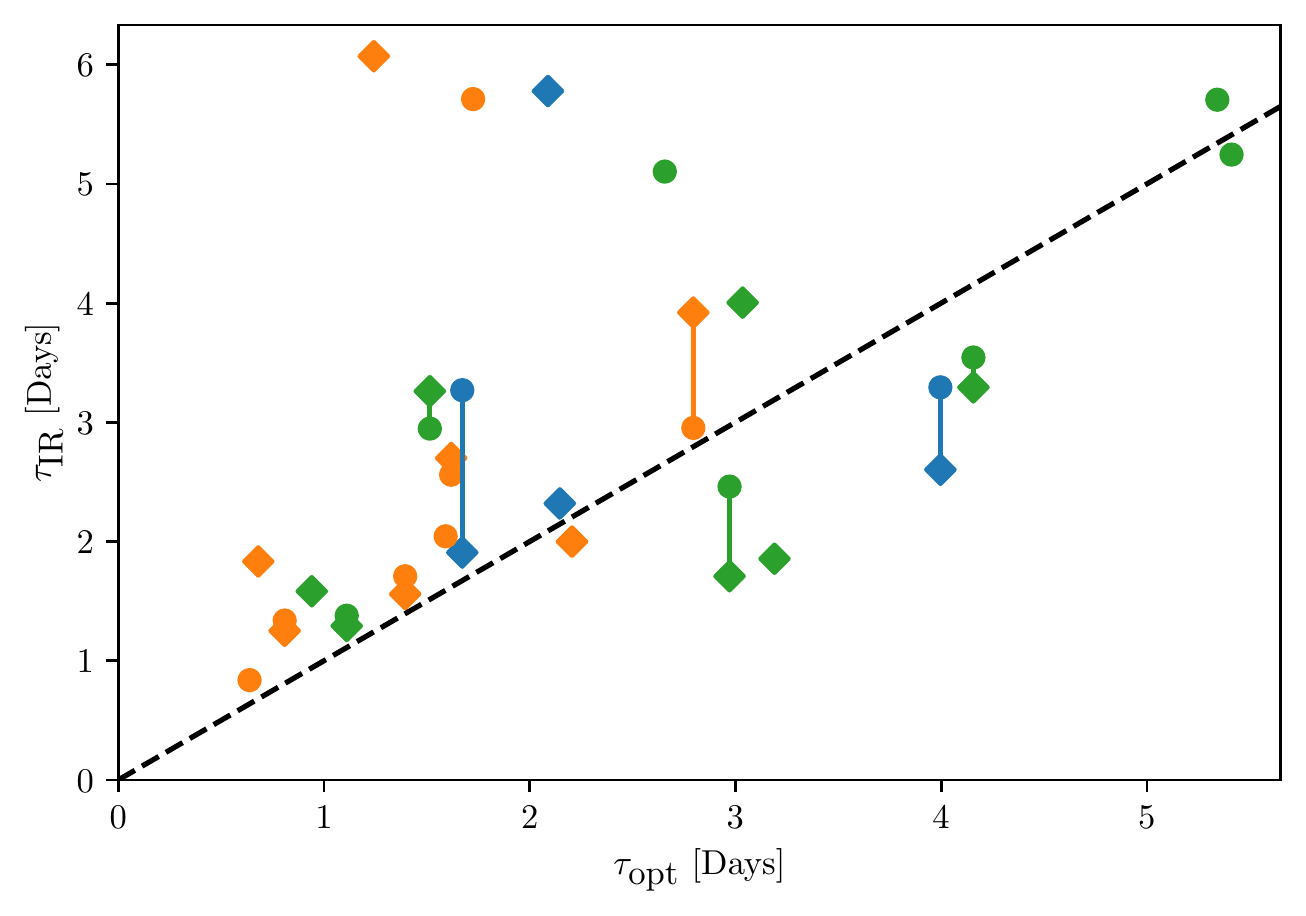}
    \caption{\edit{Comparison for the optical and infrared knee timescales for bursters (blue), dippers (orange), and symmetric variables (green). Circles show timescales from [3.6] structure functions and diamonds indicate timescales derived from [4.5] structure functions. For objects with knees in both infrared structure functions, the derived infrared timescales are joined by a vertical line. The black dashed line indicates the 1:1 correspondence locus.}}
    \label{fig:tau_comp}
\end{figure}

\edit{In Fig. \ref{fig:tau_comp} we show how the optical and infrared timescales are related. We find an overall correlation, with most of the object having a near 1:1 correspondence.
\citet{C14} perform a similar analysis and find that when periods are detected in the optical and infrared light curves, that they are similar. It is important to note the difference between our Fig. \ref{fig:tau_comp} and fig. 34 in \citet{C14}, namely that we only calculate timescales for objects with knees in both the optical and infrared structure functions, whereas \citet{C14} plot timescales for aperiodic objects. The 1:1 correlation we find suggests a common physical process for the optical and infrared behaviour.}

\section{Multi-wavelength structure functions}\label{sec: colour_sf}

Using the multi-wavelength data set at our disposal, we were able to produce time series of the \textit{CoRoT-Spitzer} colours (see Appendix \ref{sec: colour_time_series}). From these colour curves, we calculate the structure function in the same way as for the single-band light curves. Whilst \citet{Venuti21} have simultaneous \textit{K2} and VST/OmegaCAM \citep{Kuijken} observations, the latter is too sparsely sampled (with only 17 observations over the $\sim 70$ day campaign) to enable a full multi-wavelength structure function analysis.\\

The choice to calculate the structure function in magnitude space (see Section \ref{sec:sf_def}) as opposed to flux space is further justified by the fact that the colour structure function we calculate here is independent of our choice of colour (i.e., choosing to calculate the colour as \textit{Spitzer-CoRoT} rather than \textit{CoRoT-Spitzer} would yield identical results). As with the monochromatic structure functions, the value of the colour has no effect, only the relative change (i.e., we cannot tell how red or blue an object is, only how its colour varies at different timescales). Fig. \ref{fig:ex_colours1} shows the colour time series and colour structure functions for the example YSOs in Fig. \ref{fig:ex_lcs}. \edit{For comparison, we also include the respective single-waveband structure functions (right panels of Fig. \ref{fig:ex_lcs}) as faint dashed lines.}\\

\begin{figure*}
    \centering
    \includegraphics[width=\textwidth]{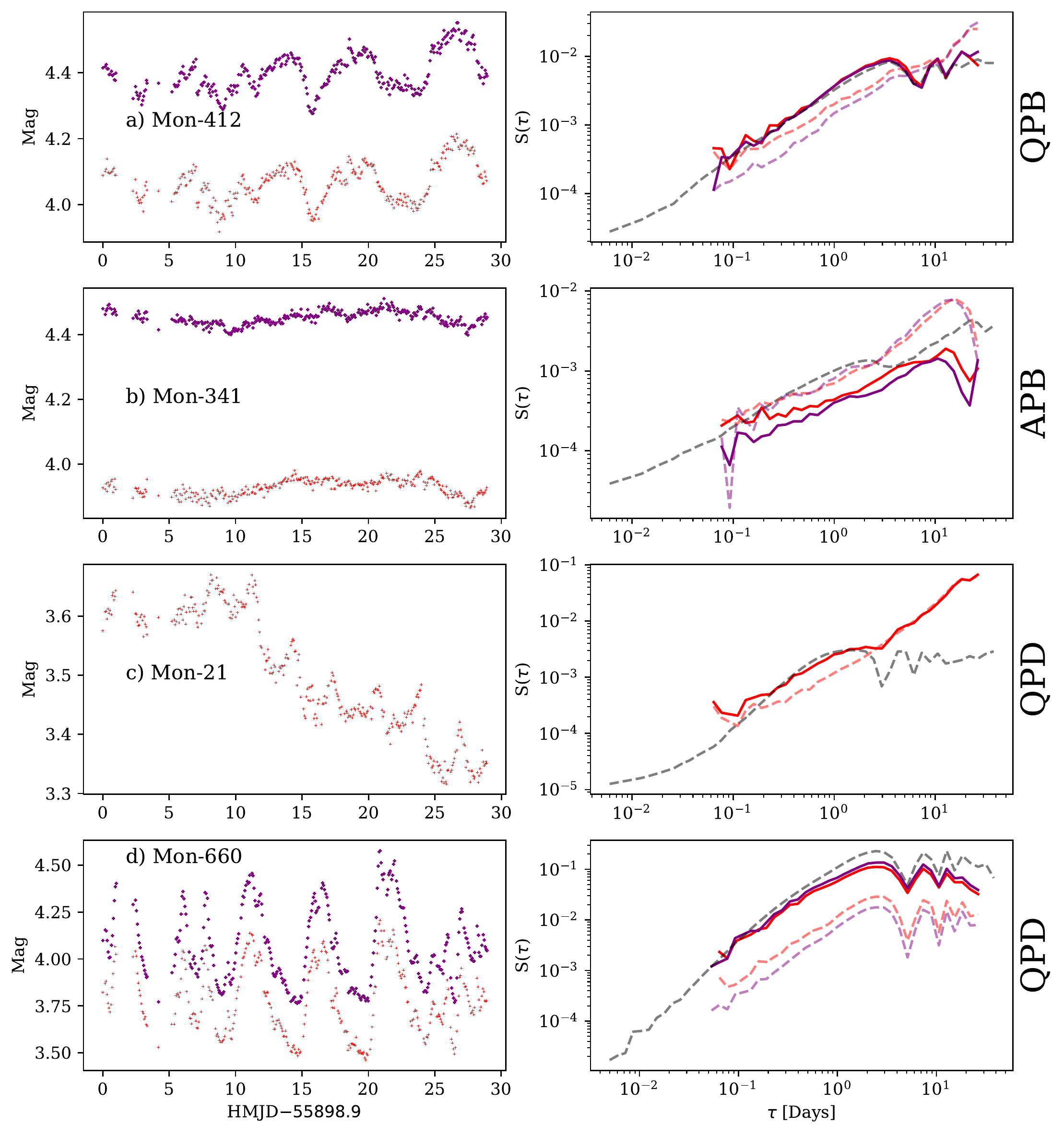}
    \caption{(Left) \textit{CoRoT} \textit{R}-[3.6] (red) and  \textit{R}-[4.5] (purple) colour time series for the YSOs in Fig. \ref{fig:ex_lcs}. (Right) The corresponding colour structure functions are shown in solid lines. For comparison, the single-waveband structure functions from the panels in Fig. \ref{fig:ex_lcs} are shown in faint dashed lines.}
    \label{fig:ex_colours1}
\end{figure*}
\begin{figure*}\ContinuedFloat
    \centering
    \includegraphics[width=\textwidth]{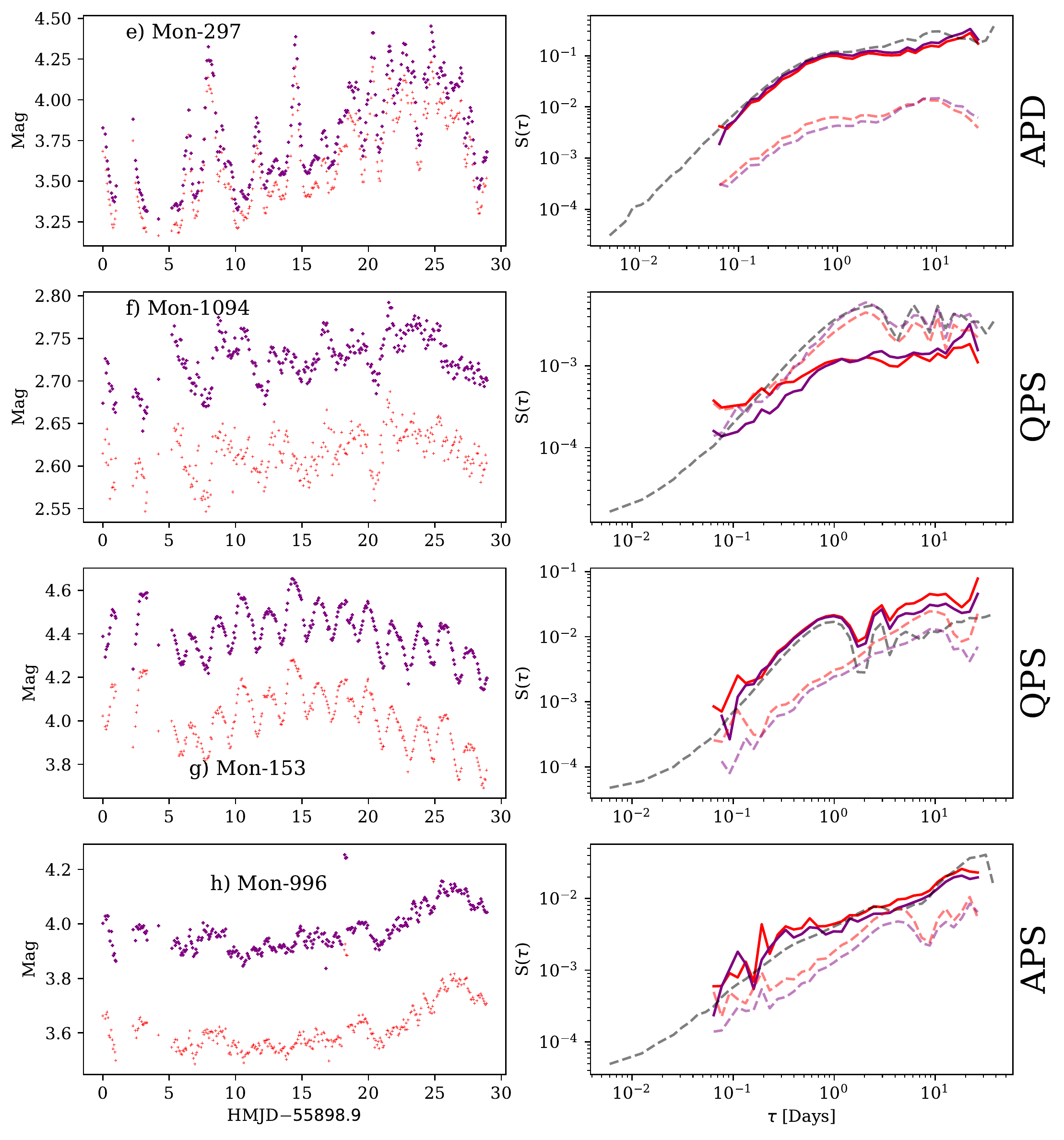}
    \caption{Continued}
    \label{fig:ex_colours2}
\end{figure*}
\begin{figure*}
    \centering
    \includegraphics[width=\linewidth]{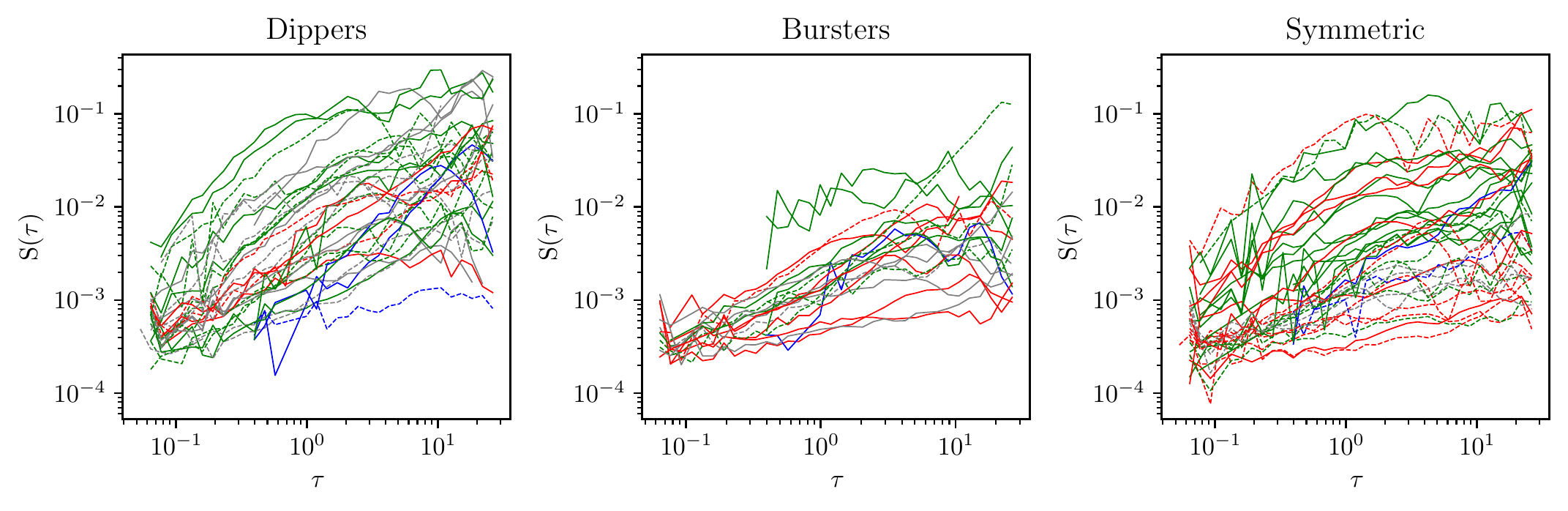}
    \caption{Structure functions of \textit{R}-[3.6] colour curves discussed in the text. Structure functions are coloured by the spectral type of the star, with blue, green, and red lines corresponding to G, K, and M stars, respectively. Dashed lines indicate quasi-periodic variables, with solid lines indicating aperiodic variables.}
    \label{fig:colour - irac1}
\end{figure*}
\begin{figure*}
    \centering
    \includegraphics[width=\linewidth]{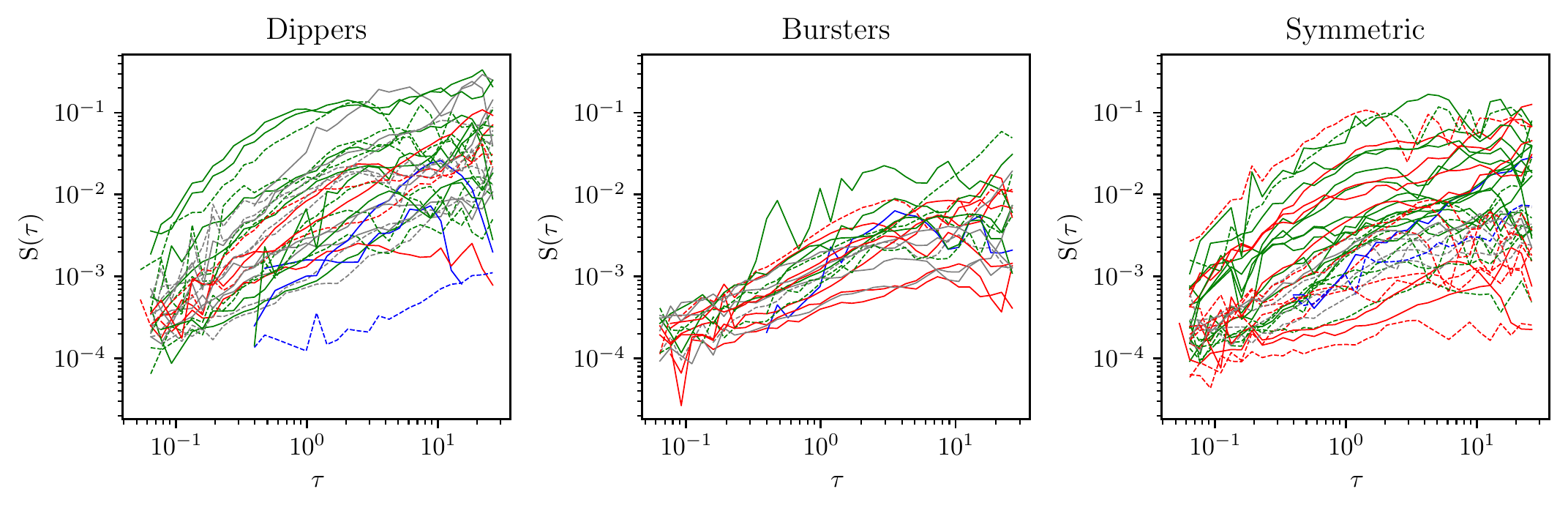}
    \caption{Structure functions of \textit{R}-[4.5] colour curves discussed in the text. Structure functions are coloured by the spectral type of the star, with blue, green, and red lines corresponding to G, K, and M stars, respectively. Dashed lines indicate quasi-periodic variables, with solid lines indicating aperiodic variables.}
    \label{fig:colour - irac2}
\end{figure*}

Figures \ref{fig:colour - irac1} and \ref{fig:colour - irac2} show the structure functions of the colour. It can be seen that the G stars, which show somewhat lower variability than their later-type counterparts in the first panel of Fig. \ref{fig:sf_corot}, display very similar colour behaviour to the K and M type dippers. In addition, the colour series for the dippers are more variable than those for the other variability types. This implies that the optical and infrared light curves are less correlated for dippers than they are for bursters. This is especially noticeable in the \textit{R}-[4.5] colour.\\

\begin{figure}
    \centering
    \includegraphics[width=\linewidth]{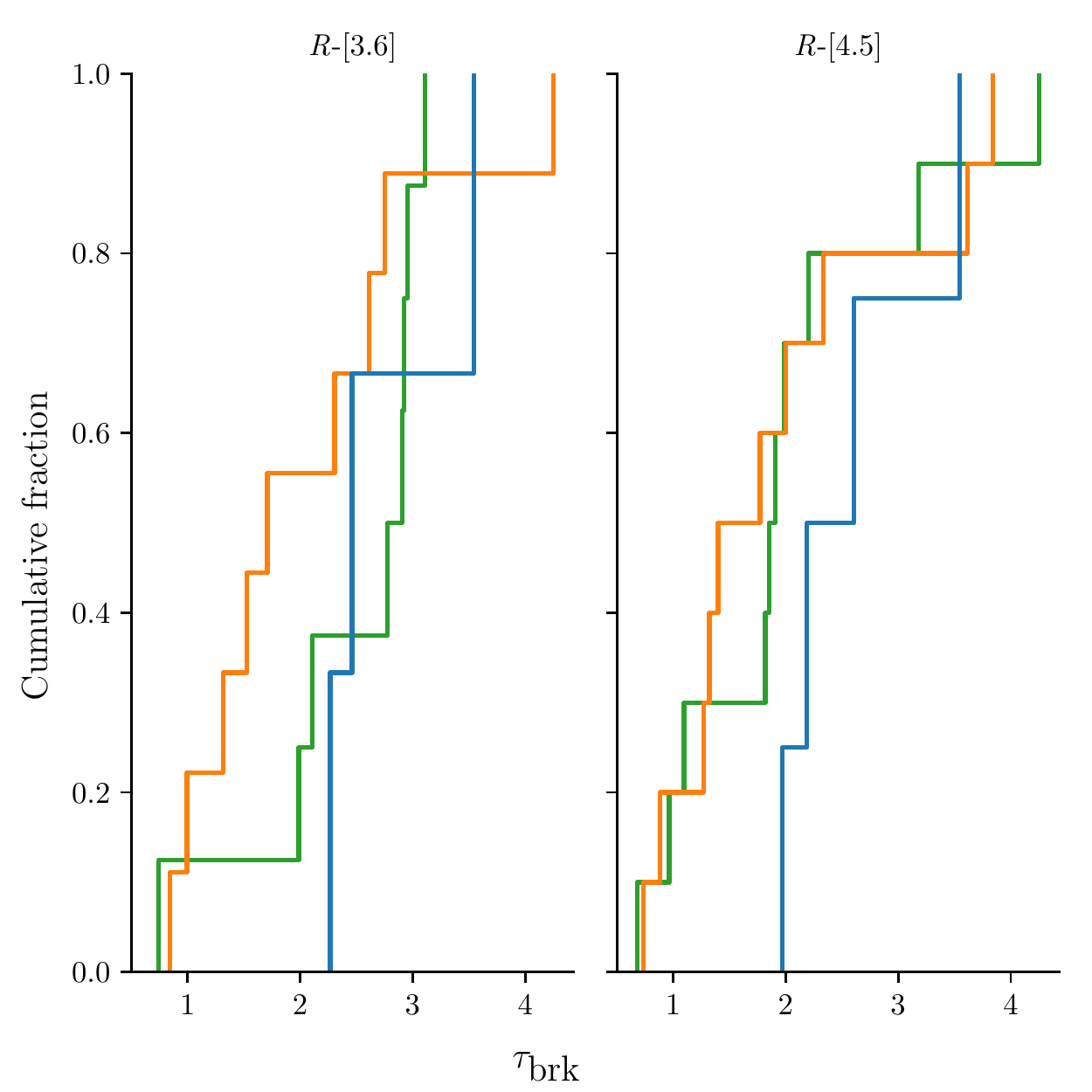}
    \caption{Distribution of timescales derived from location of knees in the colour structure functions. The bursters, dippers, and symmetric stars are represented by the blue, orange, and green distributions, respectively.}
    \label{fig:tbrk_colour}
\end{figure}

\begin{figure}
    \centering
    \includegraphics[width=\linewidth]{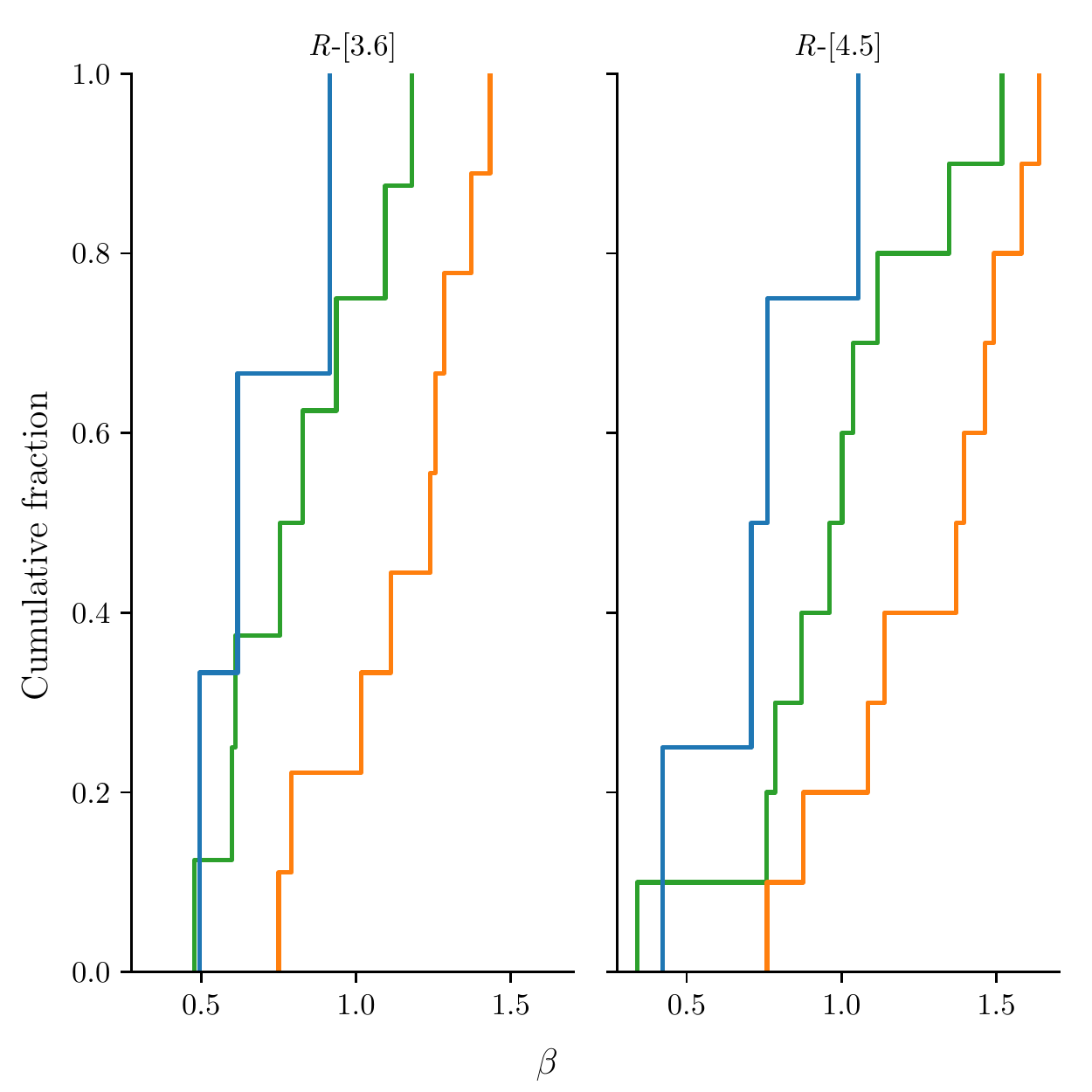}
    \caption{Distribution of the power law indices derived from least squares fitting to the colour structure functions. The bursters, dippers, and symmetric stars are represented by the blue, orange, and green distributions, respectively.}
    \label{fig:beta_colour}
\end{figure}

As in Section \ref{sec: knees}, we visually examine the structure functions to identify those objects with visible knees (see Table \ref{tab: knee_counts_ovar}). Similarly to the single wavelength structure functions, we use Equation \ref{eq:fit_eq} to extract the best fit parameters which we plot in Figs. \ref{fig:tbrk_colour} \& \ref{fig:beta_colour}. The colour timescales in both cases seem reasonably similar between the morphology classes, though our inferences are limited by the relatively small number of structure functions with knees. We also see that the dipper colour variability is much steeper than that of the bursters or symmetric variables. 

\section{Discussion}\label{sec: discussion}

\subsection{Current model} \label{sec:current_model}

\edit{Before discussing the observational findings of this paper, we wish to first summarise the current favoured model for YSO variability in order to better place our findings into context.}

\subsubsection{Dippers} \label{sec: dippers}

Optical dipping behaviour is common in YSOs, accounting for around 30 per cent of all disc bearing stars in a given region \citep[e.g.,][]{Stauffer15, Cody2018, Roggero}. It is believed that the cause of this variability is extinction by circumstellar material crossing our line of sight to the central object. There are a number of specific models to explain various light curves, for example a warped inner disc wall, caused by a tilted stellar dipole in the prototype AA Tau \citep{Bouvier}, dust trapped in accretion funnels rising above the plane of the disc \citep{Stauffer15}, or clumps of dust being produced by instabilities in the disc and levitated into our line of sight \citep{Turner}. A key observational prediction of these models is that dips should be fainter in MIR wavelengths than in optical wavelengths due to the $\sim\upmu$m sized dust particles in the circumstellar material \citep[e.g., ][]{C14}. 

\subsubsection{Bursters}

Bursting behaviour is attributed to flux from hot spots on the stellar surface caused by accretion shocks \citep{1994AJ....108.1906H}. These bursts differentiate themselves from flares caused by stellar activity by their roughly symmetric shape (i.e., the rise and decay timescales of the burst are roughly equal). For a more comprehensive discussion of current explanations for bursting behaviour, the reader is directed to section 7 of \citet{Cody17}. \edit{\citet{C14} find that the majority of the objects with strong correlation between the optical and infrared light curves are classified as bursters or symmetric variables. This is an important prediction of the accretion shock model, as both optical and infrared wavebands are in the Rayleigh-Jeans tail of the $\sim 10^6$K accretion material \citep{Calvet+Gullbring}.} We argue in Section \ref{sec: robinson} that rotational effects also strongly contribute to the bursting phenomena.

\subsubsection{Comparison with observation}

We present two pieces of evidence which support the above picture. We have shown in Section \ref{sec: colour_sf} that the variability in dippers is more chromatic than in the bursters, as expected from the variable extinction model of dipping variability. Furthermore, the shapes of the normalised magnitude distributions presented in the first panel of Fig. \ref{fig:normed_mag} show that the dippers have a much better defined maximum flux than the bursters do. We argue that this is also consistent with the current explanations. The reasoning is as follows. If the dippers are caused by occultation of circumstellar material, and the variation in continuum flux is relatively slow, there is a well defined maximum in brightness corresponding to the light from the unobscured star. By contrast, variability caused by accretion rate would presumably have a more significant bright and faint tail given that the accretion rate could increase or decrease.\\

\subsection{Light curve morphology as an effect of inclination} \label{sec: robinson}
It is commonly reasoned that dipping systems must be observed at high inclinations since the central object must be \edit{variably occulted by asymmetries in the innermost disc regions or the stellar magnetosphere}. In this section, we will explore how a similar inclination argument might explain the difference between the bursters and symmetric variables. In particular, we argue that systems observed at low inclinations are more likely to display symmetric variations, and systems at intermediate inclinations are more likely to be classified as bursters.\\

\subsubsection{Accretion variability}
\citet{Robinson} use a composite model consisting of a one-dimensional hydrodynamic simulation of the accretion column \citep{Robinson17}, non-local thermal equilibrium accretion shock \citep{Calvet+Gullbring, Robinson19}, and rotational modulation to produce synthetic optical light curves of YSOs with accretion modulated variability. They assume aligned rotational and magnetic axes, and that the in-falling material is uniformly distributed over a single hot-spot. 
\citet{Robinson} showed that by varying magnetic field properties and viewing inclination of accretion dominated YSOs, the light curves could occupy the symmetric or even dipping regions of the $Q$-$M$ plot (see Section \ref{sec: qm}). Fig. 9 of \citet{Robinson} demonstrates the general trend for the model light curves as the inclination is varied from $90^\circ$ to $0^\circ$: light curves vary from periodic to aperiodic and from purely bursting to symmetric or dipping behaviours, as characterised by the asymmetry statistic of \citet{C14} (see Section \ref{sec: qm}). The dipping behaviour is especially interesting considering the models of \citet{Robinson} do not include the effect of occulting disc material \edit{and serves to highlight the stochastic nature of accretion.}\\

\begin{figure*}
    \centering
    \includegraphics[width=\textwidth]{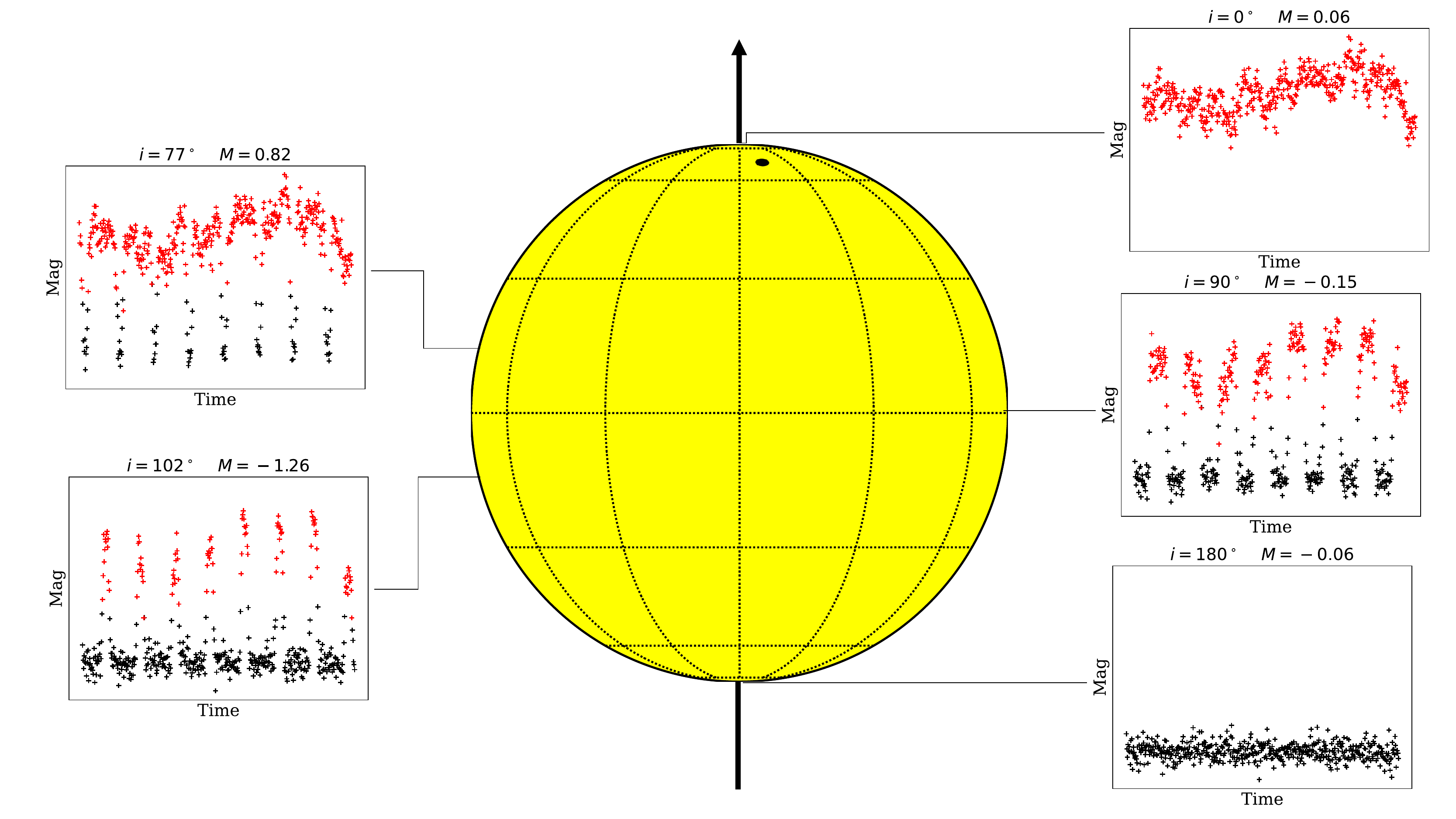}
    \caption{A schematic diagram illustrating the effect of varying the viewing angle on the asymmetry of the light curve asymmetry. The accretion hot spot is placed at a latitude of $75^\circ$. The rotation axis of the star is indicated by the vertical arrow. See text for more details.}
    \label{fig:inclination_plot}
\end{figure*}

The explanation of why changing the viewing inclination can change the morphology of the light curve is as follows. The light curve consists of a contribution from a reasonably constant photosphere, and a bright hot spot whose flux varies with the accretion rate. As the spot rotates into view, the total flux increases (and varies stochastically). So for low inclination systems, where the spot is in view for longer, the light curve is dominated by the stochastic variations driven by the variable accretion rate, with potential dips as the spot rotates out of view and we are left with just the photospheric flux. By contrast, the high inclination systems have a much smaller fraction of the rotational phase at the brighter flux level associated with the spot, and so we would describe these light curves as bursting more. Additionally, there are a number of models which would be classified as dippers despite being very low inclination models, and thus having the hot-spot visible for the entire rotational phase, highlighting the effect of the stochastic driving function. Fig. \ref{fig:inclination_plot} shows \edit{a simplified illustration of how inclination can affect light curve asymmetry}. Unlike the more models of \citet{Robinson}, we do not offer a full treatment of accretion variability. Instead we consider a constant photosphere with a single, infinitesimally small hot spot which shows stochastic variability. We place this hot spot at a latitude of $75^\circ$ and calculate the resultant light curve when viewed from a range of inclinations (note we allow $i>90$ to account for the case where we are viewing the opposite side of the star to the spot). We also add a small level of white noise to simulate observational uncertainties. To illustrate the effect of inclination alone, the accretion and white noise have identical time series for each model. The red points indicate the observations when the spot is in view and the black points indicate we only see the photosphere. Clearly the different viewing angles give rise to a range of light curve behaviours, despite the simulated accretion noise and white noise being identical between each instance. Note that, as in the simulations of \citet{Robinson}, we do not include the effect of circumstellar discs. \edit{We wish to emphasise that we are not intending this to be taken as a physical model of accretion. Instead, Fig. \ref{fig:inclination_plot} should serve simply as an illustration of how varying inclination can manifest itself as different light curve behaviour.}\\

\subsubsection{Effects of the circumstellar disc}
As previously stated, neither the schematic diagram of Fig. \ref{fig:inclination_plot} nor the more sophisticated treatment of accretion variability in \citet{Robinson} include the effects of the circumstellar disc. In this section, we briefly discuss how we expect the inclusion of the disc to affect the results of the model.\\

The first, most obvious, effect occurs when the disc intersects our line of sight to the star. In this case, we would expect the optical light curve to show dips of the form discussed in Section \ref{sec: dippers}. As we have previously discussed, these dips will be less prominent in the mid-infrared wavelengths of the \textit{Spitzer} observations. A second, more subtle, effect is that as the disc absorbs light from all directions from the star, it reaches thermal equilibrium and re-emits the light in longer wavelengths, thus acting to provide a measure of the azimuth-integrated flux from the star. Therefore by observing in infrared wavelengths, which more efficiently probe the disc flux, we can get information about the total light emitted by the star, largely unaffected by the rotational phase of the star. \edit{In principle, the signal originating in the disc should lag the stellar signal due to a combination of light-travel time and the disc thermal timescale. Radiative transfer simulations \citep[e.g.,][]{Harries} and photo-reverberation mapping studies \citep{Meng16} of circumstellar discs show that this lag occurs at a shorter timescale than our cadence allows us to sample.}\\

\edit{We demonstrate in Section \ref{sec: normed_mag} that the infrared normalised magnitudes for all variability types demonstrate similar distributions. This is perhaps unsurprising considering the large difference in wavelength. Nevertheless it seems contrary to the observations of \citet{C14}, who find that a large number of bursters and symmetric variables show strong correlation in the optical and infrared light curves. This might imply that the shape of the magnitude distributions should be reasonably independent of wavelength.}\\

\edit{We can explain this by noting that in the infrared, we are more sensitive to the light from the disc than in the optical. This implies that in the \textit{Spitzer} photometry, we are seeing the accretion variability without the rotational modulation highlighted in Fig. \ref{fig:inclination_plot}, similar to the optical light curves of the symmetric stars. Hence the bursters and symmetric variables have similar magnitude distributions in the mid-infrared wavelengths we study. Additionally, we can also see in Fig. \ref{fig:ah68_cumulative} that the overall variability of the symmetric variables and the bursters are more consistent in the infrared than in the optical. \edit{Finally, the similarity between the classes can also be seen quantitatively by performing two-sample Kolmogorov-Smirnoff tests between each pair of $\beta$ distributions. In doing this, we find that the distributions of power law index $\beta$ between the bursters and symmetric variables are considerably more similar in the infrared wavebands than the optical.} It is worth noting that these three pieces of evidence are complementary; the similarity of the magnitude distributions demonstrates that the infrared light curves explore a similar range of magnitudes, the half 16-86 inter-percentile range demonstrates that the overall level of variability is similar, and the power law indices show that the light curves have similar levels of structure.}\\

Following this logic, the infrared structure functions should probe the accretion variability. Taking this to be the case, we find the accretion to follow a power law with $\beta = 0.55 \pm 0.29$ from the [3.6] structure functions, and $\beta = 0.70 \pm 0.30$ from the [4.5] structure functions\footnote{\edit{These values of $\beta$ are derived from the combined sample of bursters and symmetric variables.}}. Note that the range quoted here is the RMS of the individual power law indices as opposed to an uncertainty for $\beta$. Although this is a reasonably large spread in $\beta$, it is important to note that this is the structure function power law index, not the more conventional Fourier spectral index (see table 6; figure 8 of \citet{S20} for comparison). The spread may also represent an intrinsic spread in the accretion profiles between different objects. The values of structure function power law indices observed here correspond to Fourier power spectra roughly comparable to a random walk (see Appendix \ref{sec: power_idx} for a discussion of this comparison). This is also consistent with the \lq long\rq\ subclass of \citet{S20}, which have values of $\beta$ varying between $\approx 0.3$ and $\approx 1.2$. This indicates that the power law relation we observe at timescales of roughly hours to weeks, may extend to timescales of years as in \citet{S20}. Whilst we include the values of $\beta$ of the entire sample, rather than just the objects without a knee in the structure function, we argue that this is a valid comparison in two ways. Firstly, there are relatively few objects with knees in the infrared structure functions, and so their effect on the overall results will be minimal. Secondly, the near-infrared data of \citet{S20} will be more sensitive to the rotation modulation from the central object than the mid-infrared data used here, requiring those authors to apply a selection to identify light curves without strong rotational effects. This is not necessary in our sample due to the longer-wavelength photometry.

\subsubsection{Threshold inclinations} \label{sec: thresholds}
Since we suspect that dipping systems are observed edge on, owing to the requirement that the \edit{optically thick circumstellar material} intersect our line of sight, we expand on the predictions of \citet{Robinson} and suggest a simple picture of the variability types, where dipping, bursting, and symmetric behaviour arises from systems observed at high, intermediate, and low inclinations, respectively. To estimate the threshold inclinations for these behaviour types, we assume that the systems in our sample have randomly distributed orientations. Therefore the fraction of systems with inclination less than a particular $i_0$ is given by
\begin{equation}
    p(i \leq i_0) = 1 - \cos i_0.
\end{equation}
We can therefore use the fraction of the sample exhibiting dipping, bursting, or symmetric behaviour to calculate the corresponding inclinations. Table \ref{tab: threshold_inclination} shows the threshold inclinations for light curves to demonstrate bursting ($i_\text{th, B}$) and dipping ($i_\text{th, D}$) behaviour in NGC 2264, $\rho$ Oph, and Upper Sco, calculated from the \citet{C14} and \citet{Cody2018} variability fractions, respectively. We also include the threshold inclinations derived from the disc-bearing young stars in the Lagoon Nebula \citep{Venuti21}. We note that, with the exception of the Lagoon Nebula sample, there is remarkable consistency between the different samples, with threshold inclinations varying by only a few degrees between the different star forming regions. \edit{\citet{McGinnis15} investigate objects in this sample which show behaviour similar to the prototype AA Tau. The inclinations derived from the extinction models in that work are all consistent with the dipping threshold in Table \ref{tab: threshold_inclination}.}

\begin{table}
\centering
  \begin{tabular}{c|c|c}
    Region & $i_\text{th, B}$ & $i_\text{th, D}$\\
    \hline
    NGC 2264 & $55^\circ$ & $69^\circ$\\
    $\rho$ Oph & $58^\circ$ & $71^\circ$\\
    Upper Sco & $54^\circ$ & $65^\circ$\\
    \hline
    Lagoon Nebula & $73^\circ$ & $80^\circ$
  \end{tabular}
  \caption{Population-derived threshold inclinations for observing bursting and dipping behaviour in the light curves of disc-bearing YSOs in a number of clusters.}
  \label{tab: threshold_inclination}
\end{table}

\subsubsection{Caveats}
There are a couple of important caveats to this picture and the inclination values quoted in Section \ref{sec: thresholds}. Firstly, in quoting these population averages, we have assumed that all the dippers are observed at the highest inclinations, whereas the simulations of \citet{Robinson} demonstrate dipping phenomena arising from low-inclination systems. Since the majority of the simulations in that paper result in either bursting or symmetric behaviour, it is reasonable to assume that most of the observed dippers are caused by line-of-sight extinction.\\

Secondly, the threshold inclinations separating the morphology types we calculate in Section \ref{sec: thresholds} are calculated from the entire population of variable \edit{Class II YSOs} in each region. The simulations of \citet{Robinson}, whilst predominantly showing the same general trend, have different threshold inclinations separating the bursters and symmetric stars, depending on the parameters of each model. Similarly, the minimum viewing angle to observe dipping will strongly depend on the geometry of the disc since we require at least part of the disc to intersect our line of sight. For example, a star with a misaligned disc \citep[e.g.,][]{2019MNRAS.484.1926D} would presumably have a different range of viewing angles which would produce dipping behaviour. We note that we have implicitly assumed that all the stars in the sample have the same threshold inclinations, and therefore that the stars have similar magnetic and disc properties. We must therefore emphasise that the values quoted in Table \ref{tab: threshold_inclination} are population-level statistics, rather than definite thresholds for each individual object.\\

A further complication is highlighted in the light curves of Mon-119 and Mon-577. These stars have symmetric light curves which show narrow dips attributed to dust in accretion columns rising above the plane of the disc. \citet{Stauffer15, Stauffer16} estimate that these objects are observed at inclinations $\lesssim 60^\circ$ owing to their lack of deep dips. \\

This picture implies that the objects categorised as dippers may also exhibit bursting phenomena, given that they lie at larger inclinations than our supposed threshold for bursting behaviour to manifest itself. As shown in Sections \ref{sec: normed_mag} and \ref{sec:sf}, the dipper light curves are generally more variable than their bursting counterparts. Therefore it is possible that, while both dipping and bursting phenomena are present, the latter is drowned out by the former. Further, there is evidence for variable accretion in many dippers \citep{Bouvier2003, Bouvier2007, McGinnis15}, indicating that even in the picture described in Section \ref{sec:current_model}, there is a significant overlap between the bursting and dipping objects.\\

\subsubsection{Summary}
We suggest a model for TTS variability dependent on viewing angle to the star. Recent simulations by \citet{Robinson} have demonstrated that inclination can have a significant effect on the optical behaviour of the system by introducing a rotational modulation. By additionally considering the effect of a disc, we can also describe the observed infrared behaviour by arguing that the disc reprocesses light from all rotational phases, leading to much more similar behaviours for bursters and dippers. Consequently, we argue that the variability in the infrared light curves of the symmetric variables and bursters directly probes the accretion variability. Following this logic, we find that the majority of burster and symmetric variables display accretion signatures roughly consistent with a random walk process.

\subsection{Comparison with \citet{Venuti21}} \label{sec: comp_venuti}

\edit{A seemingly natural comparison to draw is between this work and the work of \citet{Venuti21}, owing to the similarity in the data, and the analysis methods. However, there are significant differences between the ages and spectral types of the stars studied in these two works, and we are limited in the conclusions we can draw from any comparison of them. Nevertheless, we present here a brief comparison of the differences between our findings and those of \citet{Venuti21}.}\\

\edit{A key result of \citet{Venuti21} is the trend for stars of earlier spectral type to show systematically less variability than the later type stars. Additionally, there is a over-representation of non-variable stars in the early type sample, compared with the late type sample. Neither of these effects are observed in the sample studied here. Figs. \ref{fig:sf_corot}, \ref{fig:sf_irac1}, and \ref{fig:sf_irac2} do not show the trend of increasing variability with later spectral type found by \citet{Venuti21}. Indeed, perhaps with the exception of the \textit{CoRoT} light curves of the dippers, the spectral type of the star appears to bear no relation to the structure function. Additionally, the over representation of non-variable stars seen in the early type objects in \citet{Venuti21} is absent from the CSI 2264 data set, where the non-variable fraction of the G, K, and M stars is 13, 4, and 25 per cent, respectively.}\\

\edit{We find in section \ref{sec: sf_results} that the dippers exhibit more variability than the bursters at all but the shortest timescales. We note that this is contrary to the findings of \citet{Venuti21} who find that the maximum intrinsic variation for bursters is greater than that for dippers or symmetrically varying stars.}\\

\edit{Finally, in section \ref{sec: sf_timescale}, we find that the bursters and symmetric variables have longer median timescales than the dippers. In contrast, \citet{Venuti21} find that the dippers have the longest timescale ($\tau_\text{brk} \sim 3 $ d), with the bursters and symmetric stars having $\tau_\text{brk} \sim 1 - 2$ d.}\\

\edit{As earlier highlighted, it is perhaps unsurprising that the objects in this sample and those studied by \citet{Venuti21} show different properties, given the difference in region properties. Nevertheless we wish to draw attention to the differences in this section.}\\

\section{Conclusions}\label{sec: conclusions}

In this paper we have used the simultaneous optical and infrared photometry of 94 disc-bearing YSOs made available as part of the CSI 2264 project \citep{C14} to examine the variability properties of bursting, dipping, and symmetrically varying stars. We have analysed YSOs with variability signatures, classified by their optical light curve morphology as in \citet{C14}. Our primary conclusions are detailed below.

\subsection{Observational findings} \label{sec: conc_obs}

    We find in Section \ref{sec: normed_mag} that the magnitude distributions for the optical light curves of dippers and bursters are not simply mirror images of one another but show qualitatively different properties. By contrast, the infrared flux histograms are much more symmetric for each morphology type.\\
    
    In Section \ref{sec:sf}, we use structure functions to investigate the variability of the YSO light curves at a range of timescales. We find that the majority of the infrared light curves show increasing variability at the longest timescales accessible by the observations.  This contrasts starkly with the optical light curves which generally show a plateau (or \lq knee\rq) in their structure function, corresponding to the maximum timescale for variability. We emphasise that this is a timescale, not necessarily a period. This is notable since many objects classified as aperiodic variables by \citet{C14} show knees in their structure functions. We also find that, similar to the results of Section \ref{sec: normed_mag}, the infrared structure functions for the bursters and symmetric variables show considerable similarity, with similar distributions for the structure function power law index.\\
    
    Using the simultaneous optical and infrared data at our disposal, we perform, to the best knowledge of the authors, the first structure function analysis of colour time series data (see Section \ref{sec: colour_sf}). In doing so, we see that the variations in the bursters are significantly less chromatic than the variations in the dippers. We argue that this is consistent with the current picture of dipping and bursting behaviour, being caused by variable extinction and accretion, respectively. We note that the symmetric variables demonstrate colour behaviour which lies broadly in the middle of these two classes.

    In Section \ref{sec: normed_mag} we demonstrate that the dippers in this sample show a larger variability amplitude than the bursters and symmetric variables. In Section \ref{sec:sf}, we show this effect to be persistent at all timescales longer than a few hours. Interestingly, when \citet{Venuti21} perform a similar analysis of disc-bearing YSOs in the Lagoon nebula, they find that the bursters have a larger intrinsic variability than the dippers, and that both are more variable than the symmetric class. \edit{We wish to highlight the fact that the significant differences between the regions limits any conclusions we can draw from this comparison.}\\

    \subsection{Interpretation}
    In this section, we aim to place the findings of Section \ref{sec: conc_obs} into context by providing the interpretation of the authors. We firstly note that the findings of Section \ref{sec: colour_sf}, that the dippers have variability which is more chromatic, is consistent with \edit{the findings of \citet{C14}} the current accepted model of YSO variability (Section \ref{sec:current_model}). To this end, we also argue that the difference in optical magnitude distributions in Section \ref{sec: normed_mag} is also evidence for this model, as the dippers have a much smaller \lq bright tail\rq\  than the \lq faint tail\rq\  of the bursters. This is, we argue, consistent with the idea that the dips are caused by extinction of circumstellar material and that the variation in the bursters are caused in part by a change of accretion rate (Section \ref{sec:current_model}).\\
    
    Additionally, we suggest a model of YSO variability dependent on viewing angle to the star (see Section \ref{sec: robinson}), motivated in part by recent simulations by \citet{Robinson}. We argue that symmetric, bursting, and dipping light curves will generally arise from systems with low, intermediate, and high inclinations, respectively. This model suggests that the primary difference between the bursters and symmetric variables is the presence or absence of rotational modulation of the accretion hot spots.\\
    
    Finally, we point out the importance of the fact that the disc reprocesses light from all directions of the star. Thus by observing the system in mid-infrared wavelengths which more effectively probe the discs, one is able to negate most of the rotational modulation which separates the symmetric and bursting classes in the optical wavelengths. Invoking this model can explain why, in the mid-infrared observations, the structure functions and light curve histograms of the bursters and symmetric variables behave similarly. It also suggests that the mid-infrared variability is a direct probe of the accretion rate changes, largely free from the effects of rotation. Assuming this to be the case, we find that the power law relation for the accretion variability time series has index $\beta = 0.55 \pm 0.29$ and $\beta = 0.70 \pm 0.30$, derived from the [3.6] and [4.5] data, respectively. This is roughly consistent with the power law indices found by \citet{S20} in the \lq long\rq\  variables, which those authors also believe to represent accretion variability. Comparison between the Fourier power spectrum and structure function of correlated noise profiles implies that disc-bearing YSOs have accretion variability which roughly follows a random walk. 
\section*{Acknowledgements}

 BSL is funded by a UK Science and Technology Facilities Council (STFC) studentship. We thank Connor Robinson for helpful discussion regarding his models and Carlos Contreras Pe\~na for a careful reading of the final manuscript. This project has made use of the following Python packages: pandas \citep{pandas}, numpy \citep{2020NumPy-Array}, scipy \citep{2020SciPy-NMeth}, colorednoise\footnote{https://pypi.org/project/colorednoise/} , matplotlib \citep{hunter2007matplotlib}, and seaborn \citep{seaborn}. \edit{We are grateful to the anonymous reviewer for their thorough and insightful comments.}

\section*{Data Availability}
Data availability is not applicable to this article as no new data were created or analysed in this study.

\bibliographystyle{mnras}
\bibliography{refs.bib}

\appendix

\section{The case of Mon-804}\label{sec: mon-804}

\begin{figure*}
    \centering
    \includegraphics[width=\linewidth]{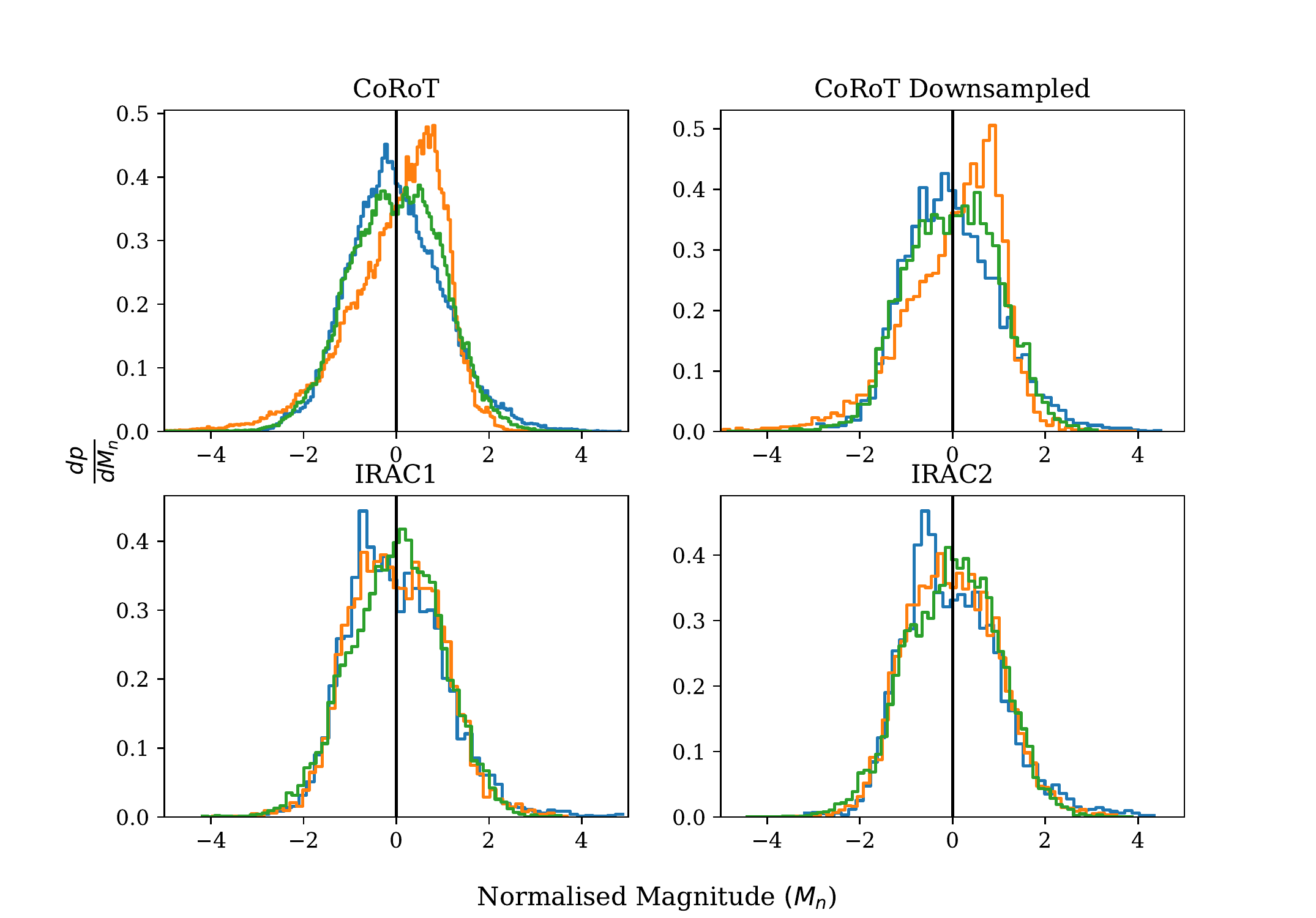}
    \caption{Normalised magnitude distributions for all objects, including Mon-804. \edittwo{As in the main text, the distribution for the bursters is represented in blue, the dippers in orange, and the symmetric variables in green.} See Section \textcolor{blue}{4} of the main text for more details.}
    \label{fig:normed_mag_with804}
\end{figure*}

\begin{figure*}
    \centering
    \includegraphics[width=\linewidth]{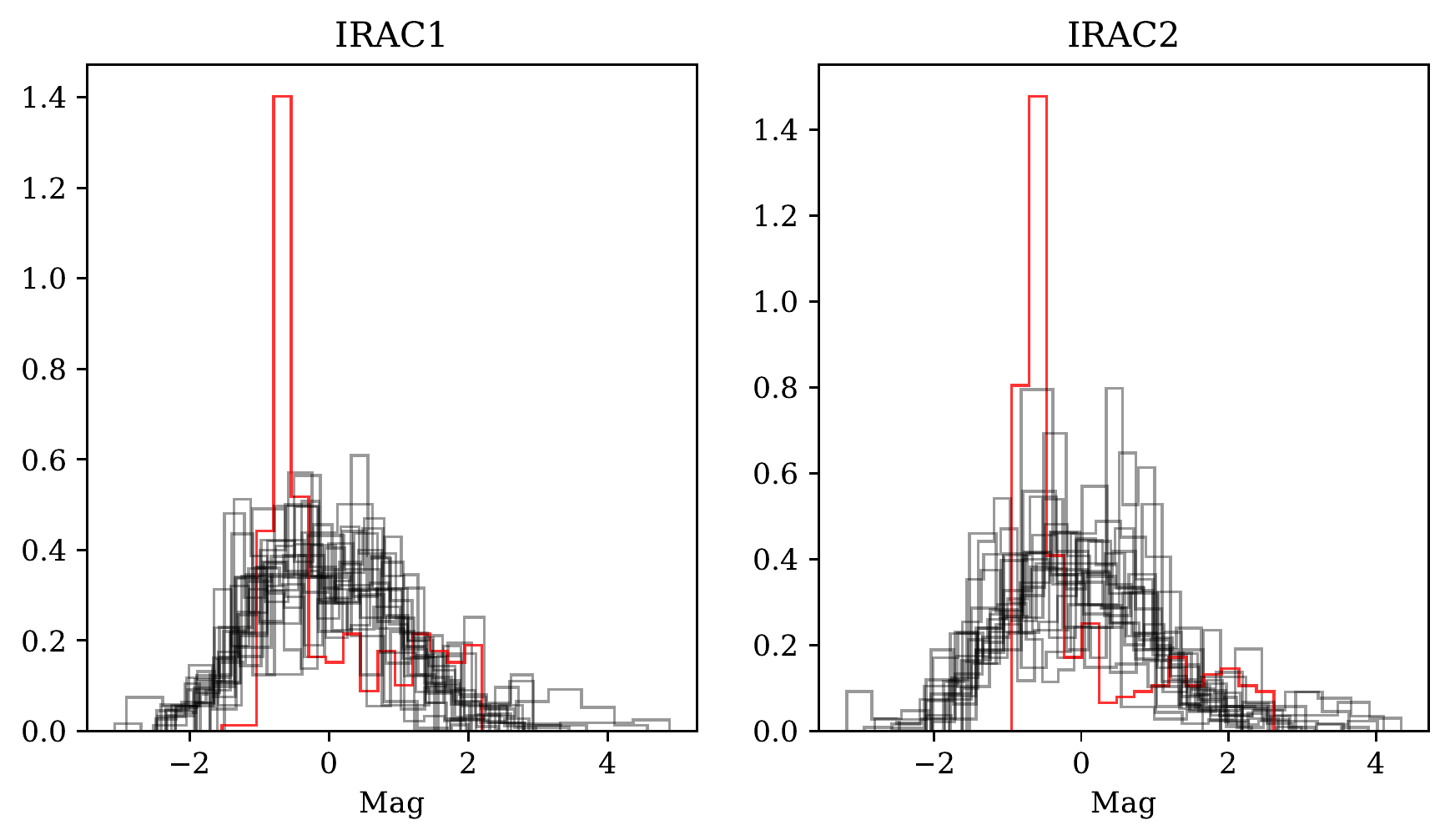}
    \caption{Individual normalised magnitude distributions for the infrared light curves of the bursters. The distribution for Mon-804 is shown in red.}
    \label{fig:mon-804}
\end{figure*}

\begin{figure*}
    \centering
    \includegraphics[width=\linewidth]{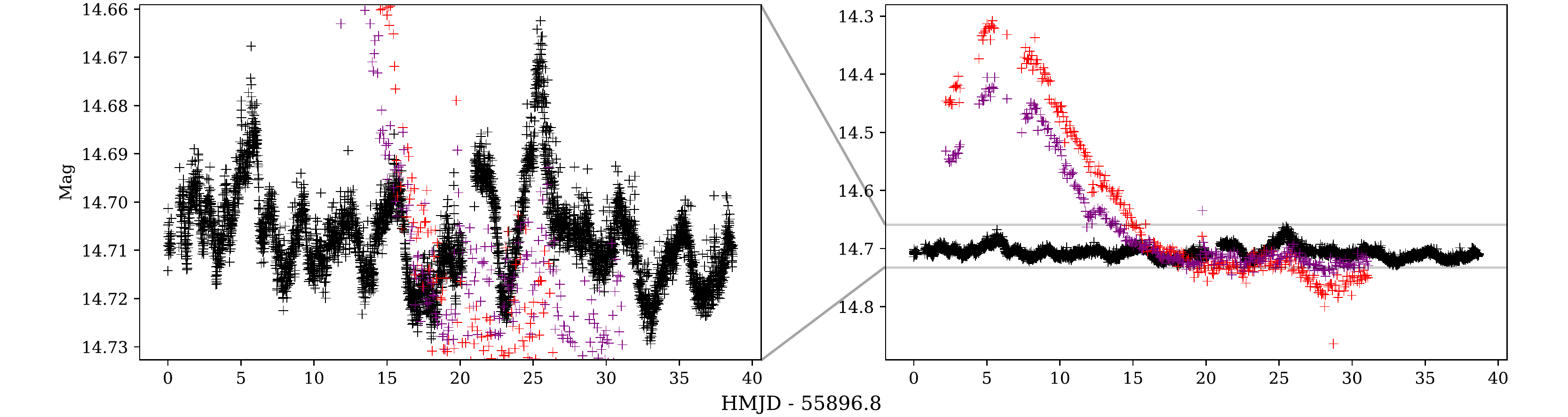}
    \caption{Optical (black) and infrared (red, purple) light curves for the object Mon-804. Owing to the dramatically different variability amplitudes of the optical and infrared light curves, we plot the light curves at the scale of the optical variability (left) and the infrared variability (right). In both cases, the infrared light curves have been shifted so that the median magnitudes overlap with the optical.}
    \label{fig:mon_804_lc}
\end{figure*}

Fig. \ref{fig:normed_mag_with804} shows the normalised magnitude distribution for all the objects in the sample, including Mon-804. There is a notable feature in both IRAC channels at a normalised magnitude of around -1.5 to -2.0. This has the effect of making the distribution appear significantly more skewed than if Mon-804 were excluded (see Fig. \ref{fig:normed_mag}). We therefore choose to exclude Mon-804 in the main work to ensure the conclusions are more representative. Fig. \ref{fig:mon-804} shows the individual normalised magnitude distributions for each burster. The red distribution corresponds to Mon-804 and highlights how it can affect the results for the entire sample. Fig. \ref{fig:mon_804_lc} shows the optical and infrared light curves of Mon-804.

\section{Colour time series} \label{sec: colour_time_series}
\edit{The \textit{R}-[3.6] and \textit{R}-[4.5] time series were produced by linearly interpolating the \textit{CoRoT} magnitudes to the timestamps of the IRAC observations. The colour time series is produced by taking the difference between the IRAC and interpolated \textit{CoRoT} measurements.\\}

\begin{figure*}
    \centering
    \includegraphics[width=0.6\linewidth]{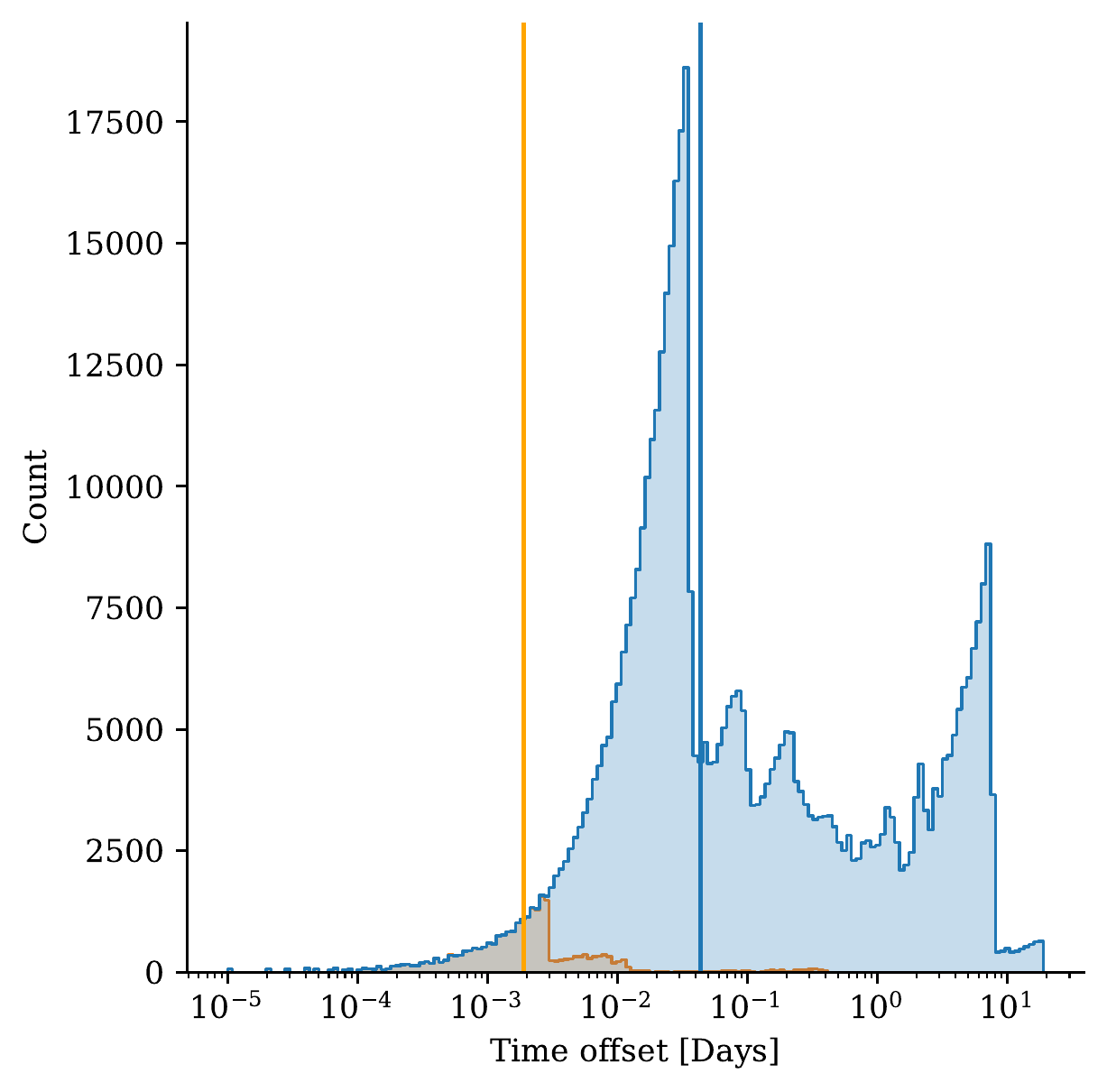}
    \caption{The distribution of time offset when finding the nearest points between bands. Blue is the offset between each \textit{CoRoT} data point and its nearest IRAC counterpart, and orange is the offset between each IRAC data point and its nearest \textit{CoRoT} counterpart. The blue and orange vertical lines mark the median offset for each scenario.}
    \label{fig:time_offset}
\end{figure*}

\edit{We followed \citet{C14} and interpolated the denser grid onto the more sparsely measured data since the effect of interpolation will be negligible. In effect, this means we interpolate between the two \textit{CoRoT} magnitudes either side of \textit{Spitzer} observation we are considering.} \edittwo{We demonstrate in Fig. \ref{fig:time_offset} that the median offset in sampling between the IRAC observations and the corresponding \textit{CoRoT} observations is around $2\times10^{-3}$ days.}

\section{Structure function power law} \label{sec: power_idx}
\begin{figure*}
    \centering
    \includegraphics[width=0.7\textwidth]{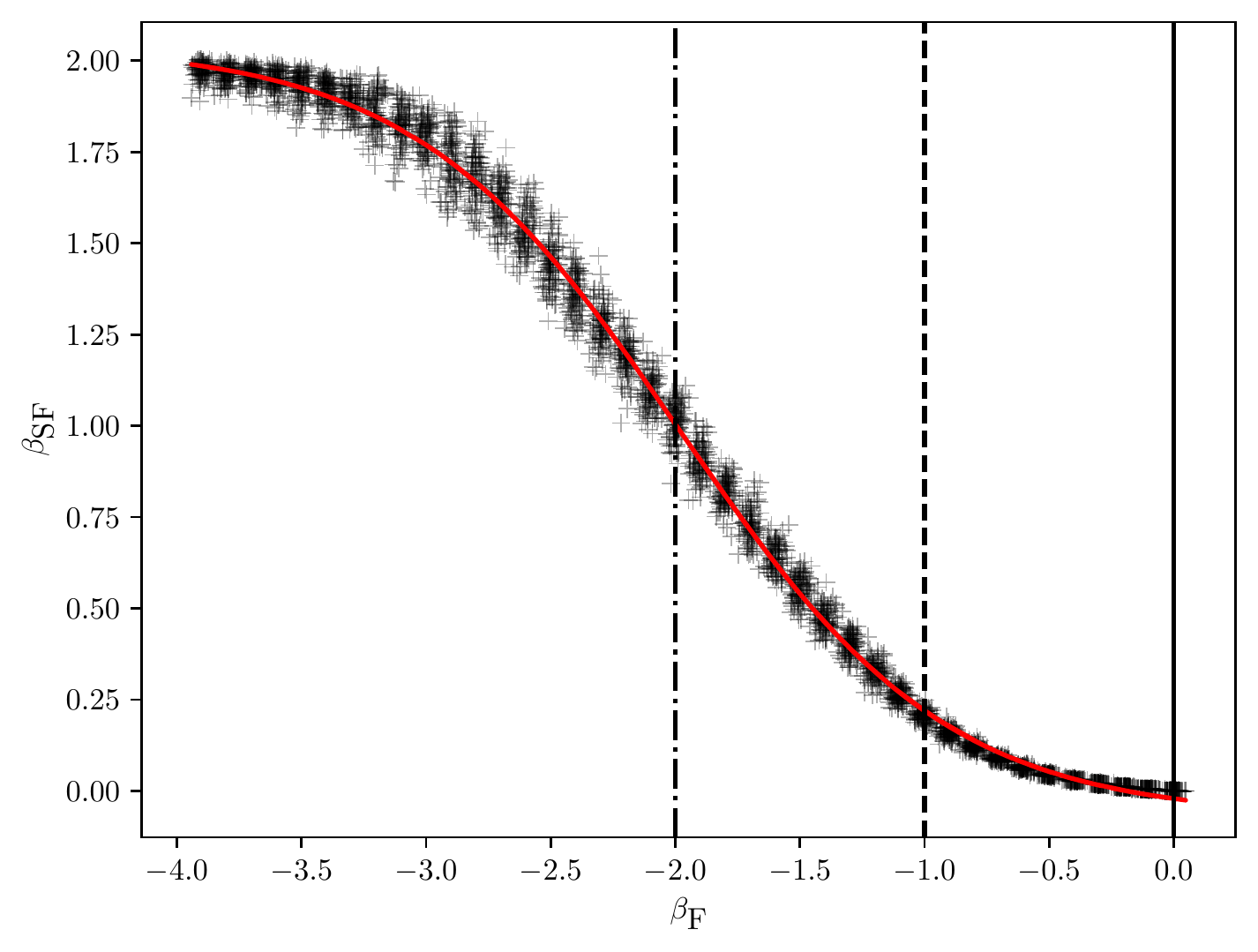}
    \caption{Relationship between the Fourier power spectrum power law index and the structure function power law index for 4000 instances of correlated noise. The best fit relation is shown in red. The solid, dashed, and dot-dashed vertical lines show the value of $\beta_\text{F}$ for white noise, flicker noise, and a random walk, respectively.}
    \label{fig:power_law}
\end{figure*}

For some particular Fourier power spectra, there exist analytic corresponding structure function power laws \citep[see][]{Findeisen}. To provide a more granular comparison between Fourier power spectra and structure functions, we simulated noise profiles using the algorithm described in \citet{1995A&A...300..707T}. We then fitted a power law to both the Fourier power spectrum and to the structure function of each noise time series. Fig. \ref{fig:power_law} shows the empirical relationship between Fourier power index and structure function index. We found a best fit relationship (shown in red in Fig. \ref{fig:power_law}) between the power law indices for the structure function and Fourier power spectrum ($\beta_\text{SF}$ and $\beta_\text{F}$, respectively) of
\begin{equation}
    \beta_\textrm{SF}  =0.99-1.06 \times\tanh\left[ {0.94\left( \beta_\textrm{F} + 1.98 \right)}\right].
\end{equation}

We do however note that there is a spread in the relation, and there is not a one-to-one correspondence between the Fourier power and structure function. We can see this in Fig. \ref{fig:sf_example}, where the structure functions for the sinusoid and the Gaussian dips have structure functions with the same power law, despite having very different Fourier spectra.

\bsp	
\label{lastpage}
\end{document}